\documentclass[graybox, envcountchap]{svmult}

\usepackage{mathptmx}        
\usepackage{amsmath}
\usepackage{amssymb}
\usepackage{color}
\usepackage{helvet}          
\usepackage{courier}         
\usepackage{dirtree}

\usepackage{makeidx}        
\usepackage{graphicx}        
\usepackage{subfig}

\usepackage{multicol}        
\usepackage[bottom]{footmisc}

\usepackage{hyperref}        
\hypersetup{colorlinks=true,urlcolor=blue}

\usepackage{ifsym}
\usepackage{tikz}
\usepackage{marvosym}

\usepackage{pdflscape}
\usepackage{booktabs}
\usepackage{ltablex}
\usepackage{url}
\usepackage{array} 
\keepXColumns          
\newcolumntype{L}{>{\scriptsize}l} 

\makeindex             

\begin{document}

\title{STOKES tables and KY codes}
\author{Jakub Podgorn\'{y} and Michal Dov\v{c}iak}
\institute{
Jakub Podgorn\'{y} (\Letter) \at Astronomical Institute of the Czech Academy of Sciences, Bo\v{c}n\'{i} II 1401/1, 14100 Praha 4, Czech Republic, \email{jakub.podgorny@asu.cas.cz}
\and 
Michal Dov\v{c}iak (\Letter) \at Astronomical Institute of the Czech Academy of Sciences, Bo\v{c}n\'{i} II 1401/1, 14100 Praha 4, Czech Republic, \email{michal.dovciak@asu.cas.cz}}
%
%
\setcounter{chapter}{7} 
\maketitle

\abstract{The family of \texttt{KY} codes is an \texttt{XSPEC}-compatible package of fully relativistic models for spectro-polarimetric modeling 
of X-ray binary systems and active galactic nuclei. Within Kerr space-time, they compute accretion disc line profiles, 
reflection spectra from discs illuminated by Comptonized coronal radiation, and thermal disc emission including disc-corona interaction and disc self-irradiation. Many of these models include polarimetric computations, and more advanced timing variants compute X-ray, UV and optical reverberation. The primary source of Comptonized X-rays is assumed to be in the lamp-post on-axis or slab above the disc geometry. The disc is presumed to be Keplerian, geometrically thin, and optically thick. The \texttt{KY} codes operate at various levels of complexity --- some models use analytical approximations for local re-processing, while others rely on detailed numerical tables. For the most accurate treatment of rest-frame re-processing in partially ionized accretion disc atmospheres, the \texttt{KY} codes employ pre-computed spectro-polarimetric tables generated with the \texttt{STOKES} code. These transmission and reflection tables are also designed for standalone use in \texttt{XSPEC} and as inputs to higher-level models beyond \texttt{KY}. The chapter first introduces the \texttt{STOKES} tables in various variants and shows their non-relativistic usage in \texttt{XSPEC} models for (i) an extended source illuminating a distant geometrically thin disc, and (ii) a compact source illuminating a vertically extended equatorial obscurer (e.g., a torus or thick wind). The relativistic \texttt{KY} ray-tracing is then described, followed by an introduction to each \texttt{KY} model. 
Examples of fits to the {\it IXPE} data of X-ray binaries and active galactic nuclei using these models are shown. Together, these models form a comprehensive and publicly available framework for spectro-polarimetric modeling and data fitting of 
accretion. 
We provide links to their online repositories.}

\begin{figure}[b]
\includegraphics[scale=.49]{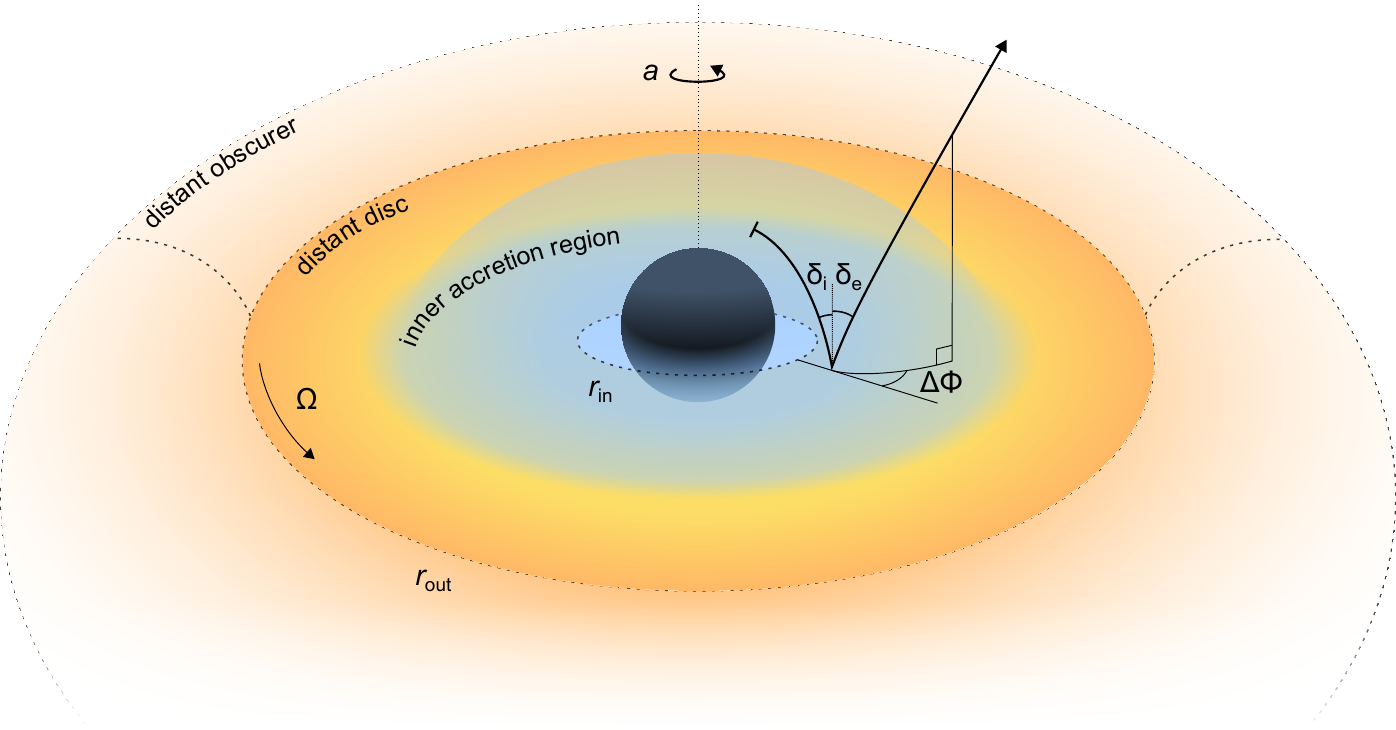}
%
\caption{\footnotesize A sketch of the geometry of accretion onto a compact object. Different flavors of the \texttt{STOKES} tables and \texttt{KY} codes families are suitable for different physical states and regions of the accreting system indicated in the illustration. The system can represent AGNs, as well as XRBs, although individual models may better adhere to one or the other. Image courtesy of Matěj Ptáček.}
\label{fig:large_scheme}       
\end{figure}

\newpage
The \texttt{KY} codes \cite{Dovciak2004, Dovciak2004b, Dovciak2008, Dovciak2011, Dovciak2014, Taverna2020, Taverna2021, Dovciak2022, Podgorny2023a, Marra2025} and \texttt{STOKES} tables \cite{Taverna2021, Podgorny2022, Ratheesh2024, Marra2025, Podgorny2025b, Podgorny2025} are X-ray spectro-polarimetric models of emission from accretion regions of X-ray binary systems (XRBs) and radio-quiet active galactic nuclei (AGNs). They are compatible with the X-ray data-fitting tool \texttt{XSPEC} \cite{Arnaud1996}, through which they can be used 
jointly for spectro-polarimetric analysis of {\it IXPE} data, or for spectroscopic analysis with instruments
such as NICER, {\it NuSTAR}, {\it Chandra}, {\it XMM-Newton} and others. This chapter aims to summarize the physics captured by the models, the main features of their outcomes, and their applicability. We refer to the \nameref{technical_docs} for links to the associated repositories where one can download each model, read the manuals and technical documentation, and find the most recent updates and literature for each sub-package. Tables \ref{tab:models1}--\ref{tab:models4} in the \nameref{technical_docs} summarize all models presented in this chapter.
\\
\\
\indent Figure \ref{fig:large_scheme} shows a sketch of an accreting system with components important for the emission and re-processing of X-rays. A black hole (BH) is shown at the center. Note, however, that some model variants can also be applied to accreting neutron stars (NS); we therefore use the more general term ``compact object'' where appropriate. We do not show jet components, because synchrotron emission and, more generally, magnetic fields are not implemented in the presented models. We will assume a prescribed coronal geometry specific to each model, and a standard Keplerian, geometrically thin, and optically thick accretion disc. Similarly, the distant components have predefined shapes and are assumed to be optically thick and static.

The \texttt{KY} codes are fully relativistic models for spectral, timing, and polarimetric signatures of the emission from the inner accretion region for a distant observer located at arbitrary inclination $i$ at spatial infinity. They all share the same general-relativistic (GR) ray-tracing kernel. Kerr space-time is assumed and photons travel globally in vacuum between the equatorial plane, the system axis, and the distant observer. The photons are emitted either with a thermal (blackbody) distribution at the equatorial disc location, or with a Comptonized (power-law) distribution at the corona location. In each case they are allowed to either directly reach the observer, or reflect from the accretion disc before arriving to the observer.
Locally, i.e. in the rest frame, photons are allowed to interact with matter, obtaining characteristic signatures of scattering, absorption, and spectral lines.  On the schematic drawing we do not suggest a precise geometry for the corona (in blue in Figure \ref{fig:large_scheme}) within the accreting system, because (a) it is still a matter of current scientific debate and (b) the codes presented adopt several geometrical prescriptions, with the aim of helping to observationally distinguish between extreme scenarios.

Each sub-model of the \texttt{KY} family reaches different levels of physical complexity. When a thermal photon from the inner disc escapes through its atmosphere or when a thermal or non-thermal photon is reflected from the atmosphere, the rest-frame re-processing is often treated using classical (semi-)analytical approximations, but in the most detailed approach, it is treated through pre-computed spectro-polarimetric transmission or reflection tables obtained with the \texttt{STOKES} code. This multi-purpose Monte Carlo (MC) simulator may be linked to standard X-ray photo-ionization codes, such as \texttt{CLOUDY} or \texttt{TITAN}, which allow for examination of a partially ionized stratified atmosphere in the non-LTE regime (see below). Although such a numerical method for re-processing in atmospheres still contains many simplifications, the pre-computed \texttt{STOKES} tables remain to date the only numerical attempt to estimate X-ray polarization emergent from a partially ionized (i.e. neither fully ionized, nor fully neutral) atmosphere. The tables can serve either as a standalone model for re-processing in the slab atmosphere
or as a local re-processing module within other higher-level codes, where they would be integrated over a pre-defined global surface geometry encapsulating large-scale accretion or ejection structures.

\bigskip
The chapter is organized as follows. Section \ref{sec:1} introduces the local re-processing \texttt{STOKES} tables, which form the basis of the more complex models presented throughout the chapter, and concludes with four non-relativistic \texttt{XSPEC}-compatible models for reflection from a \textit{distant disc} and a \textit{distant obscurer} of different shapes (see Figure \ref{fig:large_scheme}), applicable to structures such as equatorial winds, super-Eddington funnels, broad-line regions (BLRs), and dusty tori. Section \ref{sec:2} describes the relativistic \texttt{KY} codes and their use of the \texttt{STOKES} tables for modeling the \textit{inner accretion region}. 
Examples of {\it IXPE} data fitting are provided throughout both sections: in Section~\ref{sec:1}, the non-relativistic \texttt{STOKES} models are applied to the wind-obscured Galactic X-ray source Cygnus~X$-$3; in Section~\ref{sec:2}, the relativistic \texttt{KY} models are applied to the thermally dominated BH XRB LMC~X$-$3 and to the Seyfert~1 AGN NGC~4151, which includes both relativistic disc reflection and distant reflection.

\bigskip
Because in Section \ref{sec:1} we do not present any relativistic models or accretion systems that would depart from axial symmetry, we do not obtain for a distant observer any other linear polarization angle (PA) than parallel or perpendicular to the projected system axis of symmetry. Therefore, we will use the convention that the obtained linear polarization degree (PD) is positive or negative if the corresponding PA is parallel or perpendicular to the projected axis, respectively. In the relativistic \texttt{KY} models of Section \ref{sec:2}, where GR effects and asymmetric geometries allow the PA to attain arbitrary values, we follow the counter-clockwise convention of \cite{IAU1974}, measuring the PA from the projected system axis: the PA is defined as $0^\circ$ when the polarization direction is parallel to the projected system axis and increases counter-clockwise in the plane of the sky. The angle between the projected system axis and North is a free parameter in the polarimetric \texttt{KY} models, allowing direct comparison with observationally defined PAs in the IAU convention. Although the codes introduced in both Section \ref{sec:1} and \ref{sec:2} may, in principle, handle circular polarization, the physical processes implemented in them --- Compton scattering, photo-ionization, and Thomson scattering --- are not expected to produce significant circular polarization in the X-ray band for accreting compact objects. Moreover, there are currently no X-ray detectors of circular polarization (see Chapter 1), so we will omit discussing the Stokes parameter $V$ throughout this chapter.

\section{STOKES tables}
\label{sec:1}

\subsection{Radiative transfer codes used}

The local reflection and transmission pre-computations are performed using three well-established astrophysical computational codes. The ionization structure estimates are solved with radiative transfer equation solvers \texttt{TITAN} \cite{Dumont2003} and \texttt{CLOUDY} \cite{Gunasekera2025}, which do not treat the computation of polarization. The full spectro-polarimetric output is obtained using a coupled MC code \texttt{STOKES} \cite{Goosmann_2007, Marin_2012, Marin_2015, Marin_2018_UV}, which is not capable of self-consistently computing the ionization structure of the re-processing medium. Instead, it accepts the pre-computed ionization structure from \texttt{TITAN} or \texttt{CLOUDY} as input. Such a combined approach makes it possible to compute X-ray polarization from partially ionized media --- such as disc atmospheres --- where neither code alone would suffice.

\subsubsection*{TITAN}

\texttt{TITAN} \cite{Dumont2003} is an X-ray photo-ionization code, typically used for solving radiative transfer in accretion disc atmospheres, warm coronae, and warm absorbers of AGNs or XRBs. It solves the equation of radiative transfer for a plane-parallel slab in photo-ionization equilibrium (PIE), assuming energy balance and statistical equilibria for all atomic levels of the most abundant atomic species. This is the so-called non-LTE regime, which does not assume any form of local thermodynamic equilibrium. 
The non-LTE approach is essential for estimating the ionization structure of the upper atmospheric layers, which are expected to be strongly vertically stratified. These uppermost layers are the last ones the emerging X-ray radiation passes through, and therefore directly determine its spectro-polarimetric properties, for both reflected and transmitted radiation.
The code operates iteratively, using the accelerated lambda iteration (ALI) method \cite{Cannon1973, Scharmer1981, Olson1986, Hubeny2003}, originally developed in the \mbox{1970s--1980s} for stellar atmosphere calculations. The code allows illumination from both sides of the plane-parallel atmosphere.

\subsubsection*{CLOUDY}

\texttt{CLOUDY} \cite{Gunasekera2025} is a spectral synthesis and plasma simulation code designed to compute the ionization, chemical, and thermal structure of astrophysical gas and to predict its emergent spectrum from X-ray to radio wavelengths, while operating in the \text{non-LTE} regime as well. To date, it has found more diverse applications than \texttt{TITAN}, and its main focus concerns stellar atmospheres, H II regions, planetary nebulae, the interstellar and circumgalactic medium, AGN narrow-line regions and BLRs, star-forming galaxies, and irradiated exoplanet atmospheres. In its current form, the code can compute the structure of a plane-parallel slab in PIE (illumination from one side) similarly to the \texttt{TITAN} code or in collisional ionization equilibrium (CIE), assuming that the ionization state of gas is provided solely by collisions with thermal electrons at predefined temperature regardless of the source of external irradiation.

\subsubsection*{STOKES}

The \texttt{STOKES} code \cite{Goosmann_2007, Marin_2012, Marin_2015, Marin_2018_UV} is an MC simulator operating in three dimensions that injects photons from an arbitrary emission region with arbitrary spectral and predefined angular distribution into gaseous and dusty media of predefined composition, shape, time variability, and velocity. Its primary purpose is to compute polarization of the emergent photons, which are detected by an array of synthetic detectors at arbitrary inclination and azimuth at a large distance from the studied system. \texttt{STOKES} has been applied across a wide range of environments, from AGN distant components and inner regions to general astrophysical atmospheres and studies of polarization dilution by the interstellar medium. Its development began with polarized radiative transfer in the optical band, and over the past two decades it has expanded its focus to the NIR, UV and X-ray bands. In the X-rays, it accounts for multiple scattering, absorption, and spectral lines, but it does not include self-consistent free-free emission, magnetic fields, or synchrotron emission. Similarly to the \texttt{TITAN} and \texttt{CLOUDY} codes, it does not have any GR effects implemented. For the results presented in this chapter, the scattering effects are treated in the limit of cold electrons (down-scattering), although Comptonization has been recently added to the code (versions 2.34 and higher), and the \texttt{STOKES} tables and some \texttt{KY} codes have been updated for the corresponding corrections (Gupta et al., see their repositories). Preliminary results of the corrections for Comptonization can be found in \cite{Podgorny2025}.

\subsection{Local reflection tables}\label{locref}

\begin{figure}[b]
 \sidecaption
\hspace*{7mm}\includegraphics[scale=.6]{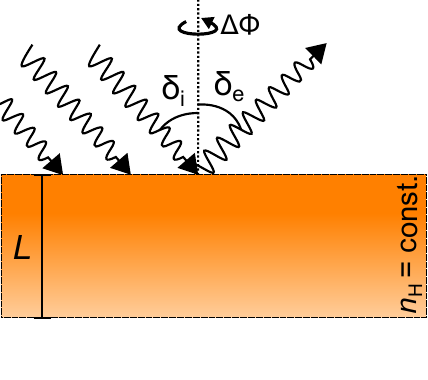}
\hspace{14mm}
\includegraphics[scale=.6]{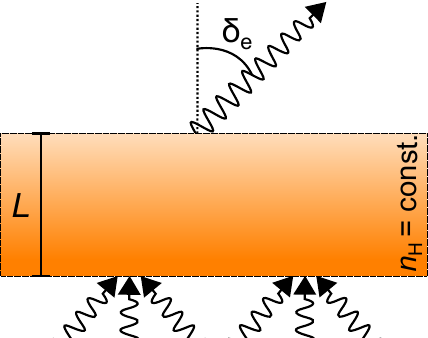}
%
\caption{\footnotesize A sketch of the local slab reflection (on the left) and transmission (on the right). Image courtesy of Matěj Ptáček.}
\label{fig:loc_ref_tr}       
\end{figure}

For the rest-frame reflection from the disc atmosphere, the \texttt{STOKES} tables assume an optically thick one-dimensional slab of constant density of neutral hydrogen, $n_\mathrm{H}$, which is depicted in Figure \ref{fig:loc_ref_tr} (left). The optical thickness is achieved through extending the physical height of the layer, $L$, to large values, so that the Thomson scattering optical depth, equivalently defined as $\tau = \sigma_\mathrm{T}\,L\,n_\mathrm{H}$ where $\sigma_\mathrm{T}$ is the Thomson scattering cross-section, reaches $\tau \sim 7$, beyond which the transmitted fraction is negligible, ensuring effectively optically thick conditions.

The constant-density assumption became in the past decades a useful approximation for X-ray spectroscopic models of disc reflection. It is also adopted in the \texttt{REFLIONX} \cite{Ross1993, Ross2005} and \texttt{XILLVER} \cite{Garcia2010, Garcia2013} local reflection models frequently used for spectral fitting of AGN and XRB sources, which, however, currently do not include polarization predictions. Various studies attempted to show the anticipated difference in both spectral lines and continuum when a full hydrostatic equilibrium with non-constant density of the atmosphere is taken into account \cite{Rozanska2002, Rozanska2011}. 
Nevertheless, quantifying the difference in spectral and polarization predictions between the constant-density and hydrostatic equilibrium cases remains to date a difficult task, owing to the computational complexity of a fully self-consistent atmospheric problem.
In realistic accretion disc atmospheres with non-constant density, the layers closest to the surface are expected to be more highly ionized due to expansion and heating \cite{Nayakshin2000, Done2007, Done2010}. Such higher ionization favors scattering over absorption processes, which would result in a lower and more energy-independent X-ray polarization fraction compared to the constant-density case.

It is only in the constant-density assumption when a useful ionization parameter, $\xi$, may be defined as \cite{Tarter1969}:
\begin{equation}\label{xi}
	\xi = \dfrac{4 \pi F_\mathrm{tot}}{n_\textrm{H}} \textrm{ ,}
\end{equation}
where $F_\mathrm{tot} = \int_0^\infty F_\textrm{E} \textrm{d}E$ is the total (energy-integrated) flux locally impinging on the slab. It represents the competition between the photo-ionization rate ($\sim n_\textrm{H}$) and recombination rate ($\sim n_\textrm{H}^2$) inside the irradiated medium. 
Broadly speaking, a higher ionization parameter corresponds to a medium more predominantly occupied by atoms in higher ionization states, and vice versa.
The polarization properties of the emergent radiation are strongly sensitive to this ionization parameter, because it determines the relative importance of scattering, absorption, and spectral line effects --- all major polarization drivers. It therefore allows us to construct a grid of numerical computations for partially ionized slab reflection, spanning from the fully neutral limit ($\xi \rightarrow 0$) to the fully ionized limit ($\xi \rightarrow \infty$), the two extremes that are more easily predicted with analytical tools (see Chapter 7). Despite the constant-density assumption and presumed solar abundance of atomic species, the ionization structure computed by a photo-ionization code such as \texttt{TITAN} is not uniform. The atmosphere is vertically stratified, with higher ionization states and higher temperature and pressure in the layers closest to the irradiating source.

The geometry of reflection is defined via three angles, depicted for the incident and reflected light ray in Figure \ref{fig:large_scheme} and from a side-view in Figure \ref{fig:loc_ref_tr} (left):
\begin{itemize}
    \item an incident inclination angle, $\delta_\mathrm{\,i}$, measured from the slab normal for an X-ray photon that is locally arriving at the slab,
    \item an emission inclination angle, $\delta_\mathrm{e}$, measured from the slab normal for a photon that escapes the slab into a particular direction,
    \item the azimuthal difference, $\mathrm{\Delta \Phi}$, which is the angle between the projected incident and emergent local directional vectors onto the slab surface plane.
\end{itemize} 
We have already discussed the sensitivity of the emergent X-ray polarization to the ionization state of the slab due to competing scattering, absorption, and spectral line effects. The polarization state of the emergent X-rays is also highly sensitive to the combination of the three geometrical angles defined above. Together, they determine the dominant single-scattering angle experienced by a significant fraction of re-processed photons --- those that are neither absorbed inside the slab nor undergo more than one scattering event. For such photons, the emergent PD can approach even 100\% when the incident and emission angles form a right angle, as follows from the general Thomson-scattering law. The same law also dictates that the prevailing PA of the emergent X-rays is perpendicular to the plane defined by the incident and emission photon momentum vectors.

\subsection{Local transmission tables}

%

The physical set-up of the local transmission problem is very similar to the local reflection problem, although the atmosphere is assumed to be of finite optical depth. As shown in Figure \ref{fig:loc_ref_tr} (right), we assume an isotropic X-ray source below the slab, which represents the thermally radiating inner layers of accretion discs of XRBs (the accretion discs of AGNs typically radiate thermally in the UV band). Again assuming constant density and solar abundance of species for simplicity, the ionization state and temperature and pressure profiles are typically inverted with respect to the vertical direction due to the opposite location of the photo-ionizing source. Together with the slab height and ionization properties, only the emission inclination angle is further needed to fully parametrize the transmission.
The outcome is not sensitive to the azimuthal angle due to isotropy of the source. Higher emission inclination angles produce higher polarization fractions
with the associated polarization angle perpendicular to the slab normal, as follows from the fundamental Thomson scattering law for polarization genesis. Absorption, which in detail depends on the ionization state of the slab, typically introduces strong energy-dependence in the X-rays and enhances the emergent polarization, because it effectively reduces the allowed scattering geometry. Photons that are not absorbed tend to escape through the shortest path to the slab surface, i.e. travelling closer to the vertical direction. As a result, their statistically most favored last scattering angle is the one formed by the slab normal and the emission inclination direction.

The above-described set-up, in which the source of radiation is located only below the slab of finite optical depth, is referred to as the Milne approximation \cite{Milne1921}. A real atmosphere of course cannot be strictly separated from the underlying heated layers and is not passive. Therefore, some X-ray polarization models \cite{Chandrasekhar1960, Loskutov1981, Ratheesh2024} assume a semi-infinite slab and/or a possibility of the source function vertically distributed inside the atmosphere according to further physical assumptions (see, e.g. Chapter 7, for the most fundamental analytically treated scenarios). Adding a source function distributed along the vertical coordinate of the atmosphere will modify the expected polarization from the Milne set-up and may even reverse the sign of the emergent PA \cite{Nagirner1962}, which also occurs for pure electron-scattering optically-thin slabs in the Milne approximation \cite{SunyaevTitarchuk1985}; in both cases due to the increasing dominance of scattering of photons travelling at large angles to the normal.

A self-consistent X-ray polarization simulation of an XRB accretion disc atmosphere remains to date computationally prohibitive. The currently available radiative transfer codes such as \texttt{TITAN}, \texttt{CLOUDY}, and \texttt{STOKES} that solve for the ionization state in the non-LTE regime and take into account scattering, absorption, and spectral lines, are suitable for the Milne approximation, which we will assume further. A limitation is that the optical depth $\tau$ of the slab remains a free model parameter --- increasing it arbitrarily enhances the energy-dependent scattering/absorption effects on the escaping radiation, while the Milne approximation simultaneously assumes no photon emission within the slab, which becomes increasingly unphysical for large $\tau$.

A further limitation of the above-described set-up for XRBs is that disc atmospheres in the inner accretion region are illuminated from both sides --- from below by thermal radiation from the deeper disc layers, and from above by non-thermal (coronal) radiation or by thermal radiation returning to the disc. In reality, the transmission and reflection problems therefore cannot be fully separated, and their combination is necessary to reproduce observed spectra and polarization. Although \texttt{TITAN} and \texttt{STOKES} are capable of double illumination of a slab, the two scenarios are treated separately in the models presented further, which is a computationally simpler and more flexible approach. The observed X-ray polarization from a more realistic double-illuminated XRB atmosphere would be lower and less energy-dependent than in the results presented further, because the atmosphere will be, on average, more ionized and less absorbing due to the presence of the second source. More precise evaluation, however, remains to be computed.

\subsection{Non-relativistic models for \texttt{XSPEC}}

\subsubsection*{Reflection from a disc atmosphere}

\begin{figure}[b]
\includegraphics[scale=.55]{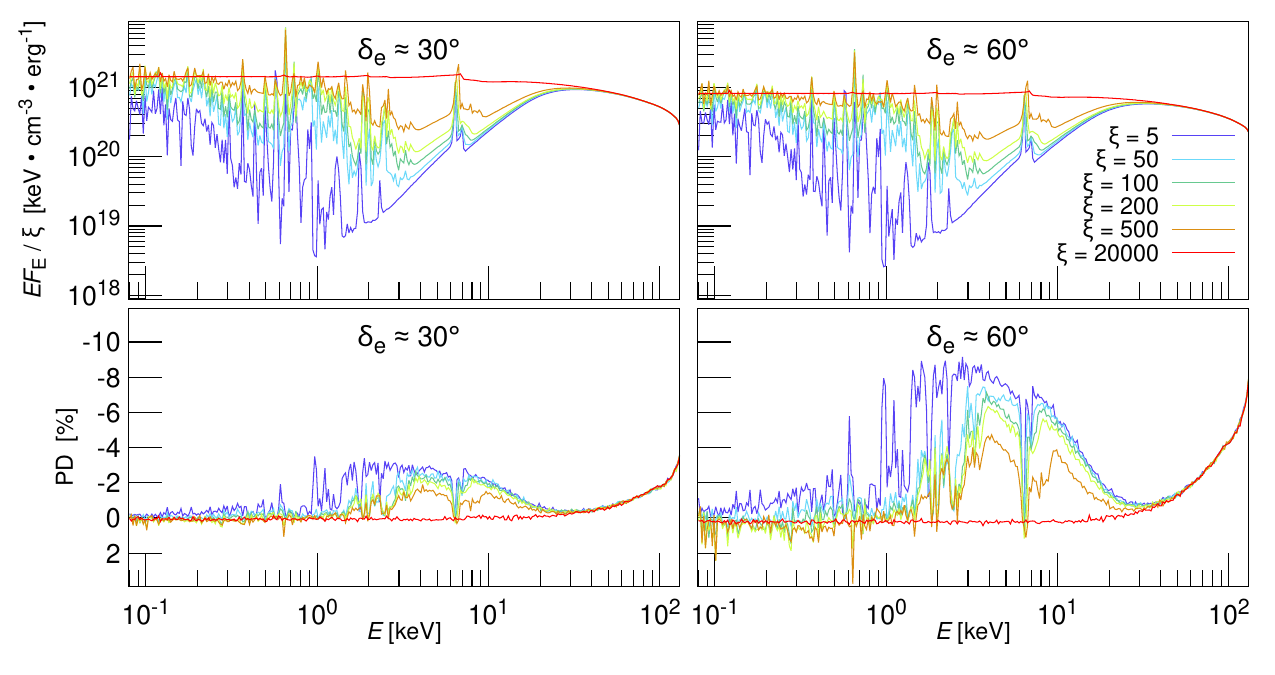}
%
\caption{\footnotesize The spectra (top row) and the polarization degree vs. X-ray energy (bottom row) locally emergent after re-processing in a constant-density slab. We show results for an inclination of $30^\circ$ (left column) and $60^\circ$ (right column). The polarization degree is rather low, because the results are integrated in emergent azimuthal angles and the unpolarized illumination is isotropic in intensity. The color code represents different values of the ionization parameter in [$\textrm{erg} \,\,\textrm{cm} \,\,\textrm{s}^{-1}$]. Low ionization (blue) shows (less polarized) spectral lines and (more polarized) continuum absorption features. High ionization (red) converges to scattering-induced energy-independent polarization degree and the re-processed spectrum mirrors the incident power-law radiation. The spectra are shown as $EF_\mathrm{E}/\xi$, i.e., they are corrected for the slope of primary power-law radiation ($\mathrm{\Gamma} = 2$) and amplitude from flux normalization, so that we can better differentiate subtle spectral features. Image adapted from \cite{Podgorny2023a}.}
\label{fig:results_loc_reflection_xi}       
\end{figure}

We will first introduce the local reflection tables for power-law reflection from a disc atmosphere in PIE \cite{Podgorny2022} that were computed using the \texttt{TITAN} and \texttt{STOKES} codes combined. Both codes were run in the identical geometrical set-up described in Section \ref{locref}. The table parameters are the three angles $\delta_\mathrm{\,i}$, $\delta_\mathrm{e}$, and $\mathrm{\Delta \Phi}$, the ionization parameter $\xi$, and the power-law index $\mathrm{\Gamma}$ of the incident radiation. They are computed for three independent incident polarization states, which allows for interpolation of arbitrary incident polarization state. The emergent Stokes parameters $I$, $Q$, and $U$ are computed in 300 energy bins logarithmically spaced between 0.1 and 100 keV, an energy range common to all \texttt{STOKES} table variants described in this chapter. Such resolution is sufficient for polarization analysis of, e.g., {\it IXPE} observations. The slab was vertically stratified in 50 layers, i.e. sufficiently for the approximation of the optically thick scenario.

The spectral part of the tables was compared against the \texttt{REFLIONX} and \texttt{XILLVER} tables, showing a reasonable agreement in both normalization and shape despite the dramatically different method of computation \cite{Podgorny2022}. Although estimation of X-ray polarization from reflection off accretion discs has a long history \cite{Chandrasekhar1960, Matt1993, Poutanen1996b}, the polarization properties of the re-processed radiation with respect to $\xi$ are a unique property of the \texttt{STOKES} tables. They currently represent the sole numerical attempt to compute X-ray polarization of reflection from a partially ionized optically thick slab in PIE. The tables were compared to the fully neutral and fully ionized limits with corresponding (semi-)analytical prescriptions, showing a reasonable agreement \cite{Podgorny2022, Podgorny2025b}. Another unique property of the tables is the ability to separate the thermalized soft X-ray reflection sub-component from the scattered sub-component dominating in the hard X-rays, made possible by the coupled radiative-transfer and MC approach in which the thermalization process can be switched off in the MC component. This makes the \texttt{STOKES} tables potentially useful for disc-corona energy exchange and reverberation studies.

Figure \ref{fig:results_loc_reflection_xi} shows a selection of the tabulated spectra and PD versus energy, integrated in the \{$\delta_{\mathrm{\,i}},\mathrm{\Delta \Phi}$\} angular space. Different values of the ionization parameter are shown in the color code. The spectra share typical X-ray reflection features, such as increased presence of absorption and spectral lines for low $\xi$. In polarization, the spectral lines are preserved as dips due to intrinsically unpolarized state of dominating fluorescent line emission, which, however, can still be scattered while escaping the slab, so the polarization at energies of the (broadened) fluorescent lines is not zero and can have a different PA from the continuum. The Compton hump is a dip in X-ray polarization due to the increased contribution of multiple scatterings at those energies. The energy-dependent photo-electric absorption reduces the average number of scatterings, which narrows the range of scattering geometries and induces polarization, similarly to the transmission problem. For the lowest $\xi$ (in dark blue in Figure \ref{fig:results_loc_reflection_xi}), the values between 3--5 keV already reach the analytical predictions with single-scattering approximations inherent to fully neutral reflection. The angle-averaged display in Figure \ref{fig:results_loc_reflection_xi}, which is convenient for inspecting spectral features in detail, shows that the predicted local polarization fraction increases with inclination. However, the integration over incident inclination angles (as isotropic intensity illumination with Simpson weighting) and azimuthal angles dilutes the polarization fraction compared to individual directional table cells, where a specific combination of incident and emission ray directions defines a dominant scattering angle and can produce significantly higher polarization fractions. The way we integrate in the angular space dictates the predominantly negative total PD, i.e., PA perpendicular to the disc's normal.

\begin{figure}[b]
\begin{tikzpicture}[%
x=1pt, y=1pt,          
inner sep=0pt,         
outer sep=0pt]

\node[anchor=south west] (base) at (0,0)
  {\includegraphics[width=\linewidth,     
                   trim={3.6cm 1.2cm 5.0cm 0.3cm},
                   clip]{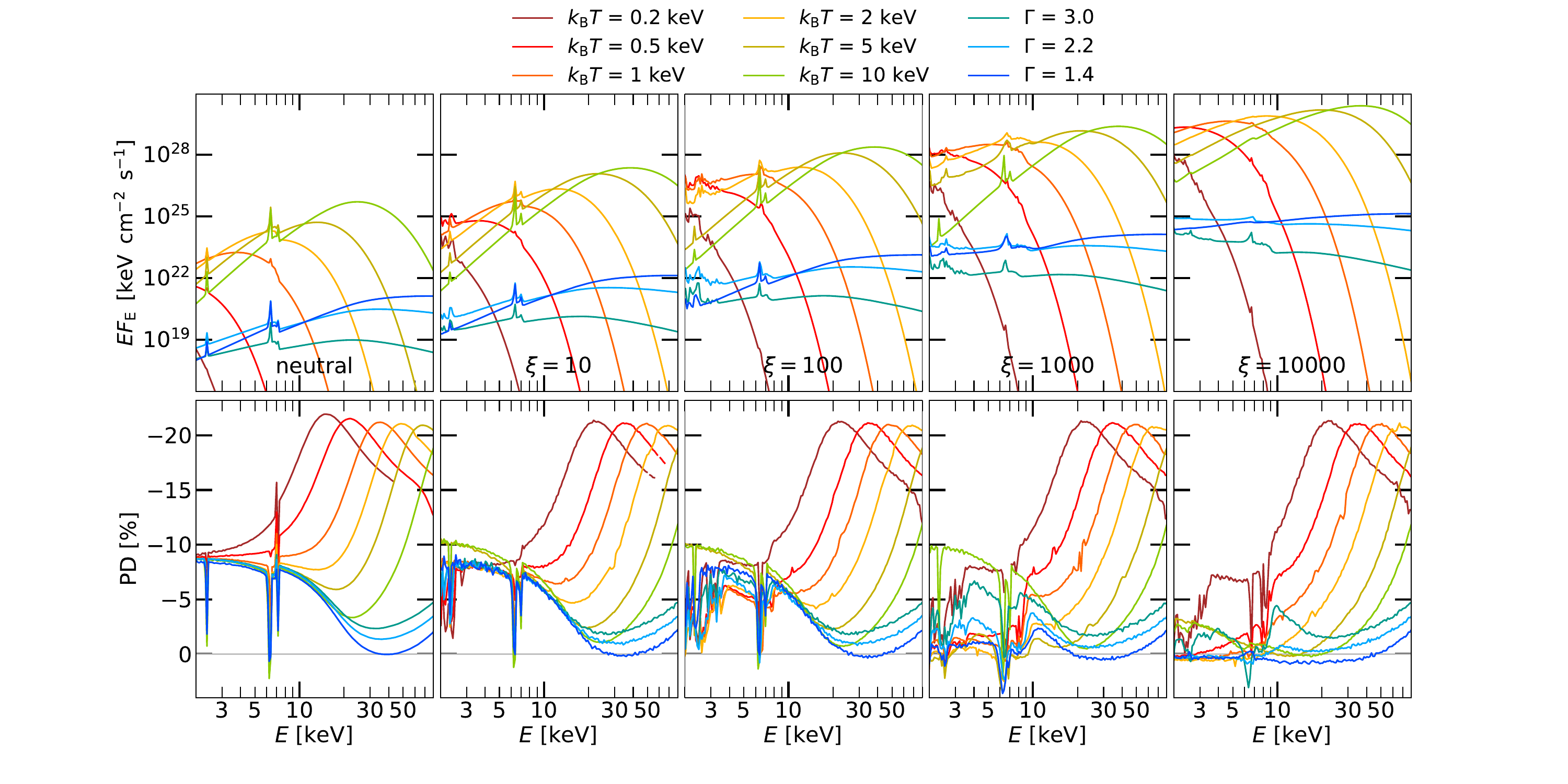}};
%
\node[anchor=south west] at (83.5,27)
  {\includegraphics[width=0.095\linewidth]{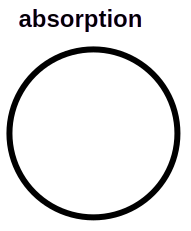}};

\node[anchor=south west] at (151,49)
  {\includegraphics[width=0.09\linewidth]{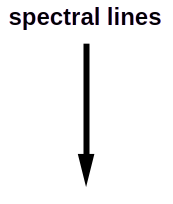}};

\node[anchor=south west] at (95,23)
  {\includegraphics[width=0.12\linewidth]{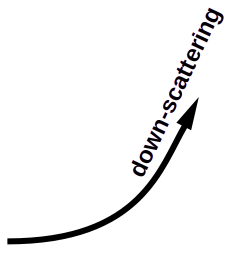}};

\end{tikzpicture}
\caption{\footnotesize The spectra (top row) and the PD (bottom row) vs. X-ray energy locally emergent after re-processing in a constant-density slab. The results are integrated in all incident inclination angles (using a Simpson rule for isotropic incident intensity), all emission azimuthal angles, and plotted for one emission inclination angle of $60^\circ$. In the color code, we show different examples of blackbody temperatures, $k_\mathrm{B}T$, for incident thermal emission and power-law indices, $\mathrm{\Gamma}$, for incident Comptonized emission. The shape of the incident spectrum is reflected in the energy dependence of the scattering-induced polarization, particularly towards higher energies, due to down-scattering effects in the shown geometry. The amount of absorption and spectral line features decreases in both spectra and polarization with increasing ionization parameter (from left to right columns in [$\textrm{erg} \,\,\textrm{cm} \,\,\textrm{s}^{-1}$]). The left-most panel shows reflection from fully neutral matter with normalization for $\xi = 1\,\, \textrm{erg} \,\, \textrm{cm}\,\, \textrm{s}^{-1}$ as described in \cite{Podgorny2025b} for the power-law reflection, and analogously for blackbody reflection. Image adapted from \cite{Podgorny2025}.}
\label{fig:results_loc_reflection_BB}       
\end{figure}
More recently, a variant of the reflection tables, called \texttt{STOKESBB} tables, for illumination with a single-temperature blackbody spectral energy distribution was computed \cite{Podgorny2025}. The table parameter $\mathrm{\Gamma}$, characterizing the incident radiation, is in this case replaced by the temperature of the blackbody incident radiation, $k_\mathrm{B}T$. Figure \ref{fig:results_loc_reflection_BB} shows the reflected spectra and PD versus energy for six values of $k_\mathrm{B}T$, compared to three examples of the previously described power-law illumination with $\mathrm{\Gamma}$ in the color code. The shape of the incident spectrum is reflected in the energy dependence of the scattering-induced polarization, particularly towards higher energies for certain geometries, due to Compton down-scattering effects. The most energetic incident photons are down-scattered to lower energies, so that photons observed at lower energies have undergone more scatterings on average than those observed at higher energies. Multiple scatterings---more likely to occur for incident rays perpendicular to the slab plane---tend to mix the scattering planes and broaden the distribution of scattering angles of individual scattering events, which, due to the vectorially additive nature of polarization, reduces the net polarization. The scattering-induced continuum polarization fraction therefore increases with energy, if the slab is illuminated preferentially from a direction parallel to the normal of the slab (such as for the isotropic intensity illumination law). In the mid and soft X-rays, scattering produces energy-independent results due to fast convergence to the elastic limit of Thomson scattering. On top of scattered radiation, we see the same imprints of spectral lines and photo-electric absorption with respect to $\xi$, which increases from left to right panels, as in Figure \ref{fig:results_loc_reflection_xi}. Both \texttt{STOKES} and \texttt{STOKESBB} tables have their variants for fully neutral reflection, shown in the left-most panel of Figure \ref{fig:results_loc_reflection_BB}.

\subsubsection*{Transmission through a disc atmosphere}

\begin{figure}[h]
 \sidecaption
\includegraphics[width=0.54\linewidth]{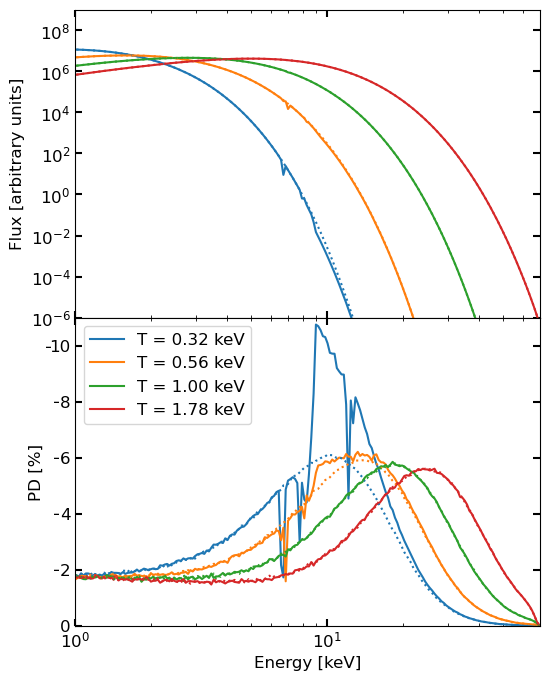}
%
\caption{\footnotesize The locally emergent spectra (top) and polarization fraction (bottom) vs. energy after transmission through a highly ionized homogeneous slab with $\tau = 0.67$ in PIE for $\xi = 10^6 \,\,\textrm{erg}\,\, \textrm{cm} \,\,\textrm{s}^{-1}$ viewed under $75^\circ$ inclination. In different colors, we show different blackbody temperatures. The solid lines show the full simulation, the dotted lines show the scattering-only results without the contamination by absorption and spectral lines. The peak in polarization fraction shifts with the temperature of the incident radiation due to down-scattering effects on cold electrons, similarly to reflection from a cold slab. Image adapted from \cite{Marra2025}.}
\label{fig:trans_BB}       
\end{figure}

The most comprehensive grid of local slab transmission computations in PIE and CIE with the effects of scattering, absorption, and spectral lines was performed in \cite{Marra2025} with \texttt{CLOUDY} and \texttt{STOKES} for the purpose of relativistic disc integration inside the \texttt{KY} codes. The parameters of the corresponding \texttt{STOKESBBTRANS} tables are $\delta_\mathrm{e}$, $\tau$, $k_\mathrm{B}T$ of the incident single-temperature blackbody, and $\xi$ in PIE or the slab temperature, $T_\mathrm{slab}$, in CIE. Figure \ref{fig:trans_BB} shows examples of the emergent spectra and polarization fraction versus energy for different temperatures of the incident blackbody spectrum in PIE. The corresponding PA is orthogonal to the slab normal, as expected for slabs of moderate $\tau$ in the Milne approximation. We see that the down-scattering effect on emergent polarization fraction versus energy, discussed with Figure \ref{fig:results_loc_reflection_BB}, applies to the transmission problem as well. However, the assumption of cold electrons relative to the inner disc radiation is debatable, and inelastic scattering off warm electrons — as in warm coronae — would counter-act the down-scattering effect on emergent polarization. If the electron temperature equals that of the incident blackbody, Compton scattering predictions for polarization match well those of elastic scattering. On top of scattering features, spectral lines manifest as dips in polarization and the energy-dependent absorption opacity manifests in the reversed ionization edge near 9 keV for the blue curve with $k_\mathrm{B}T = 0.32$ keV.
\begin{figure}[h]
 \sidecaption
\includegraphics[width=0.54\linewidth]{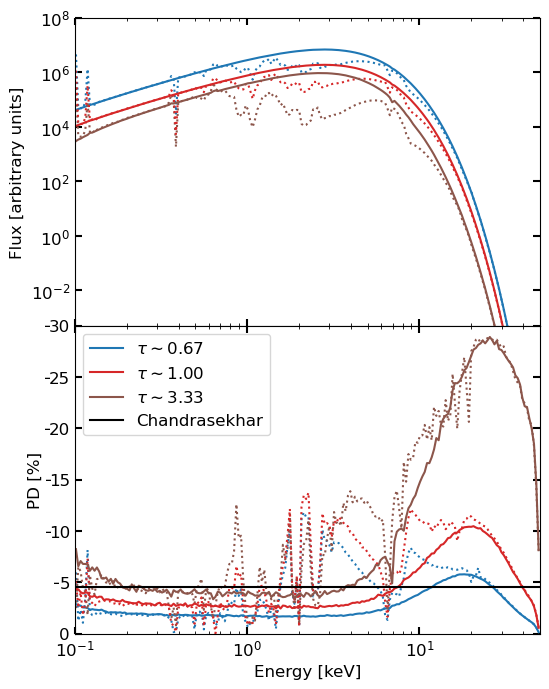}
%
\caption{\footnotesize The locally emergent spectra (top) and polarization fraction (bottom) vs. energy after transmission through a homogeneous slab in PIE (solid lines) and CIE (dotted lines) viewed under $75^\circ$ inclination. The results are shown for a 1 keV temperature of the blackbody incident radiation, and for three different optical depths, $\tau$, of the slab in colors. For high $\tau$, the energy-independent scattering-induced plateau in polarization degree near 0.2--2 keV converges to the Chandrasekhar's prescription for semi-infinite electron-scattering atmosphere, which is shown in black. Image adapted from \cite{Marra2025}.}
\label{fig:trans_tau}       
\end{figure}

Figure \ref{fig:trans_tau} shows the same quantities, but compares the PIE and CIE regimes and different $\tau$ values for one selected blackbody temperature and inclination. The CIE regime generally results in a less ionized slab than PIE, the latter producing smoother spectra and polarization profiles with energy, less affected by absorption and spectral lines. The same ionization structure can in principle be obtained under various physical conditions and regimes. But regardless of how it is sustained, it is the final ionization state that determines the X-ray polarization profile of the emergent radiation through the relative contributions of scattering, absorption, and spectral lines inside the slab. For the atmospheric transmission cases shown in this chapter, the obtained slab stratification was negligible; hence, they were produced with a single slab layer of variable height $L$.

The remaining parameters $\tau$ and $\delta_\mathrm{e}$ alter the magnitude of PD, but roughly preserve its energy profile and the PA (if $\tau \gtrsim 0.4$ in the Milne approximation). Increasing $\tau$ increases polarization of the emergent radiation due to increased absorption and down-scattering in our assumption of a cold slab. Both processes reduce the number of scatterings for emergent photons per observed energy; the former at energies with strong photo-electric absorption opacity contribution, the latter at hard X-ray energies. The pure Thomson scattering limit of the PD (clearly visible in PIE between 0.2 keV and 4 keV in Figure \ref{fig:trans_tau}) converges for $\tau \gtrsim 4$ to the Chandrasekhar limit for electron-scattering semi-infinite atmospheres \cite{Chandrasekhar1960}, shown as a black line in Figure \ref{fig:trans_tau}. Higher/lower inclination increases/decreases the emergent polarization fraction, as predicted by classical analytical results for limiting cases (see Chapter 7) due to the general increase/decrease of asymmetry in the source-slab-observer system with inclination.

\subsubsection*{Extended source reflecting from a geometrically thin slab}

\begin{figure}[b]
 \sidecaption
\includegraphics[scale=.35,     
                   trim={1.cm 4.2cm 11.0cm 2.3cm},
                   clip]{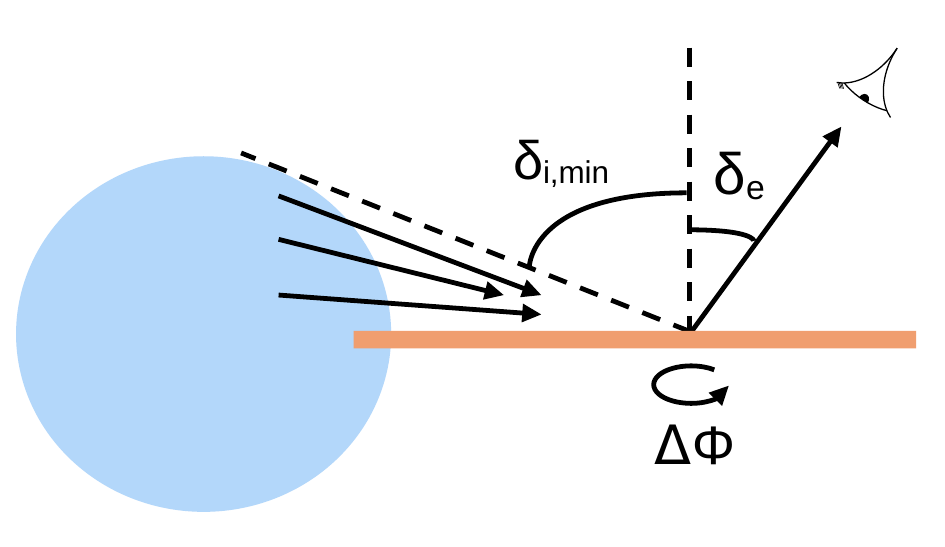}
%
\caption{\footnotesize A sketch of the reflection geometry assumed by the \texttt{STOKES\_DISC} model. A central vertically extended source of Comptonized emission is illuminating the outer disc in the equatorial plane. Image adapted from \cite{Podgorny2024}.}
\label{fig:xsstokes_disc_sketch}       
\end{figure}
\begin{figure}[b]
%
\begin{tikzpicture}[%
x=1pt, y=1pt,          
inner sep=0pt,         
outer sep=0pt]

\node[anchor=south west] (base) at (0,0)
  {\includegraphics[width=\linewidth]{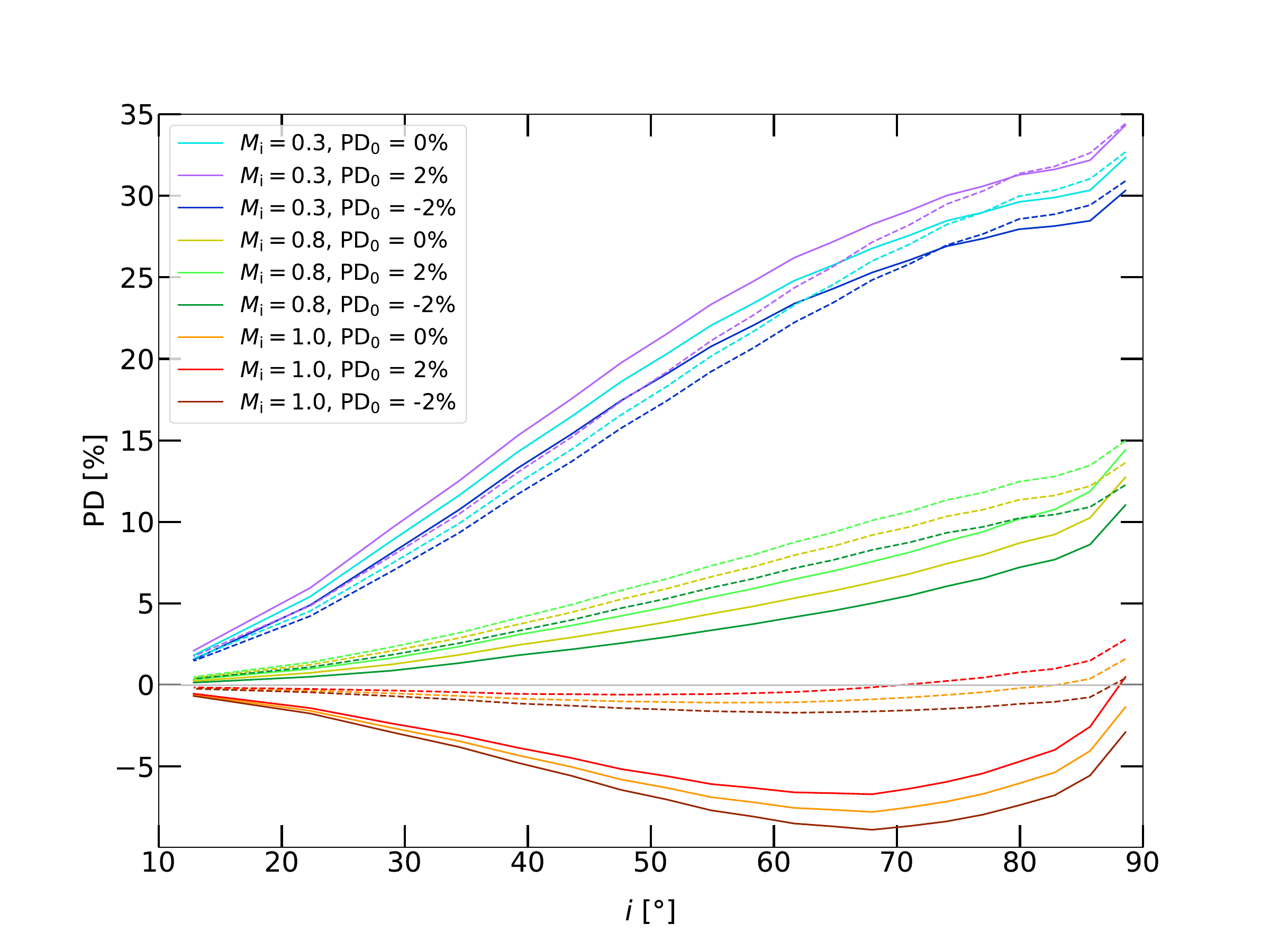}};
%
\node[anchor=south west] at (52,102)
  {\includegraphics[width=0.037\linewidth]{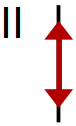}};

\node[anchor=south west] at (52,39)
  {\includegraphics[width=0.05\linewidth]{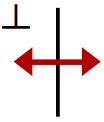}};


\end{tikzpicture}
\caption{\footnotesize The polarization fraction vs. inclination of the observer, $i$ (or locally $\delta_\mathrm{e}$), predicted by the \texttt{STOKES\_DISC} model in 3.5--6 keV (solid lines) and 30--60 keV (dashed lines). The different combination of colors show results for different geometrical extensions of the corona $M_\mathrm{i} = 0.3$ (top), $M_\mathrm{i} = 0.8$ (middle) and $M_\mathrm{i} = 1$ (bottom). Each triplet of lines shows the response to changing incident polarization for the cases of an unpolarized source, a source polarized at 
2\% parallel to the disc axis, and a source polarized 2\% perpendicular to the disc axis. Image adapted from \cite{Podgorny2024}.}
\label{fig:disc_mue_dependent}       
\end{figure}

The \texttt{STOKES\_DISC} model \cite{Podgorny2024} is a non-relativistic \texttt{XSPEC}-compatible model for distant disc reflection that makes use of the \texttt{STOKES} power-law reflection tables. It can be used additively in \texttt{XSPEC} alongside dedicated models of the inner accretion region, absorbers, or other cold non-relativistic reflectors. The model integrates the nearly neutral part of the reflection tables over a range of incident inclination angles (using the Simpson and trapezoidal integration rule for isotropic incident intensity over the range) and all azimuthal angles (uniformly), assuming a geometrically thin distant disc residing in the equatorial plane. The set-up represents a distant accretion disc illuminated by a radially and vertically extended central corona, and may be applied to hot inner-accretion flow scenarios and ADAF accretion solutions typical for low-luminosity AGNs (LLAGNs). Figure \ref{fig:xsstokes_disc_sketch} shows the presumed model geometry. The corona is indicated in blue but its shape is arbitrary --- the model only assumes a given solid angle subtended by the source as seen from the reflecting point on the far-away disc. In the meridional plane, the opening angle of the vertically extended corona is defined by the minimum incident inclination, $\delta_\mathrm{\,i,min}$, above which the corona illuminates the distant disc isotropically in intensity. The observer views the equatorial disc plane under inclination $i$, which coincides with the local emission inclination angle, $\delta_\mathrm{e}$, of the \texttt{STOKES} reflection tables used.

The inclination together with the coronal size, represented by a dimensionless parameter $M_\mathrm{i} = \cos{\delta_\mathrm{\,i,min}} \in [0;1]$, are the primary parameters determining the global reflection geometry and thus the observed polarization properties. Figure \ref{fig:disc_mue_dependent} shows the dependence of PD on inclination and the parameter $M_\mathrm{i}$, integrated in two energy bands without the presence of prominent spectral lines. Higher inclination again typically increases the observed polarization fraction. The coronal size $M_\mathrm{i}$ scales the PD magnitude and may even change the sign. The highest positive\footnote{Throughout this chapter, a positive/negative PD denotes a PA parallel/perpendicular to the projected system axis of symmetry, respectively.} polarization is achieved for a system viewed edge-on when the corona has the lowest height, because in such a case, the allowed angular space for emission and scattering directions is significantly reduced and the overall symmetry is suppressed: the predominant single-scattering planes (for nearly neutral reflection) of individual photons coincide with the equatorial plane and a PA parallel to the disc axis then emerges. The other extreme is a complete isotropic illumination of an extended corona covering the entire reflecting disc ($M_\mathrm{i}\rightarrow1$) that favors photons arriving vertically over those arriving radially, which results in net negative PD of generally lower magnitude than for a compact central corona due to increased isotropy in the global scattering geometry. The profile of PD with energy is provided by the underlying local reflection tables (see Figure \ref{fig:results_loc_reflection_xi}). The dependence on incident coronal polarization and energy is also depicted in Figure \ref{fig:disc_mue_dependent}, but the sensitivity to these parameters is not as high as to $i$ and $M_\mathrm{i}$, which determine the large-scale reflection geometry. As a practical application, the \texttt{STOKES\_DISC} model has been used to estimate X-ray polarization from distant disc reflection in LLAGN 2110 observed by {\it IXPE} \cite{Chakraborty2025}.

\subsubsection*{Compact source reflecting from a vertically extended obscurer}

\begin{figure}[b]
 \sidecaption
\includegraphics[scale=.26]{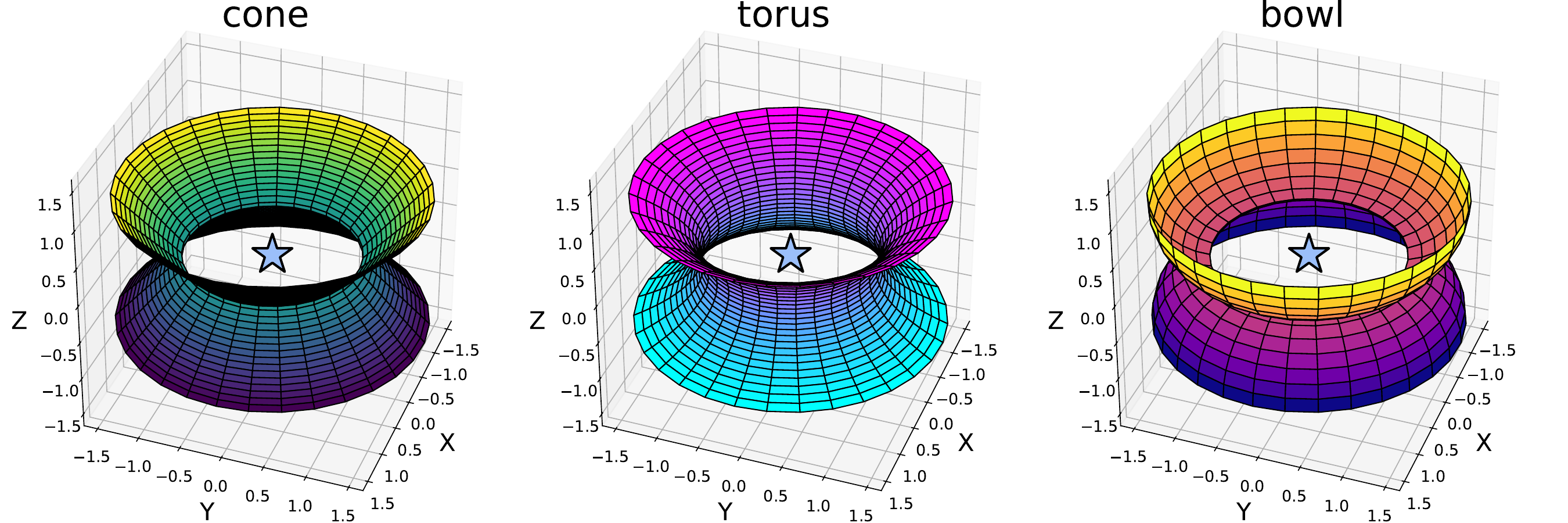}
%
\caption{\footnotesize A sketch of the three shapes studied for the elevated reflecting equatorial obscurer, as viewed by an inclined observer. Image adapted from \cite{Podgorny2025b}.}
\label{fig:sketch_3D}       
\end{figure}
\begin{figure}[b]
 \sidecaption
\includegraphics[scale=.44,     
                   trim={3.cm 4.7cm 8.0cm 3.cm},
                   clip]{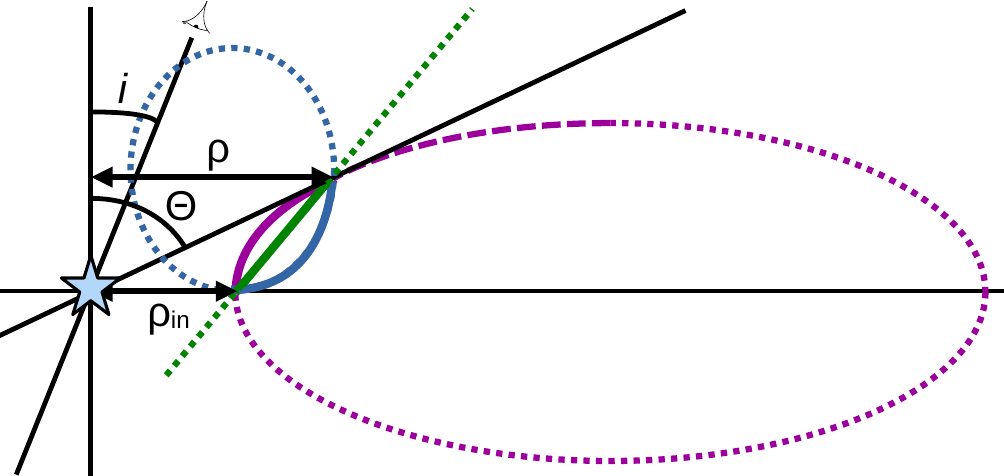}
%
\caption{\footnotesize Parametrization of the three shapes studied for the elevated reflecting equatorial obscurer, shown for one meridional quadrant. The reflecting inner walls are rotationally symmetric around the system axis and mirrored across the equatorial plane. Image adapted from \cite{Podgorny2025b}.}
\label{fig:geometry_sketch}       
\end{figure}
The \texttt{STOKES\_CONE}, \texttt{STOKES\_TORUS}, and \texttt{STOKES\_BOWL} models \cite{Podgorny2024, Podgorny2025b} are non-relativistic \texttt{XSPEC}-compatible models for reflection from a vertically extended optically thick equatorial obscurer, making use of the \texttt{STOKES} reflection tables. Each model represents a different predefined shape of the reflecting surface: a cone, an elliptical torus, and a bowl. Each can be combined additively with other physical component \texttt{XSPEC} models to fit data from an entire XRB or AGN. The models integrate the local reflection tables across the inner walls of the obscurer, approximating the central source as point-like compared to the size of the distant reflector. The set-up, depicted and parametrized in Figures \ref{fig:sketch_3D} and \ref{fig:geometry_sketch}, may represent large-scale reflection from thick winds or super-Eddington funnel structures in both XRBs and AGNs, and Compton-thick dusty tori or BLRs in AGNs. The power-law central X-ray source can be a first-order approximation of a NS boundary/spreading layer or a BH corona. To accommodate its diverse applicability, each model allows the ionization profile of the reflector and the spectro-polarimetric and anisotropy properties of the source to be freely set. The main geometrical parameters are the observer inclination, $i$, the opening angle, $\mathrm{\Theta}$, and the skew, $\rho/\rho_\mathrm{in}$, of the inner walls. Each of the three models has a variant with a single-temperature blackbody source instead of the power-law source, using the \texttt{STOKESBB} reflection tables. Analogically, these are called \texttt{STOKESBB\_CONE}, \texttt{STOKESBB\_TORUS}, and \texttt{STOKESBB\_BOWL} models.

\begin{figure}[t]
 \sidecaption
\includegraphics[scale=.49]{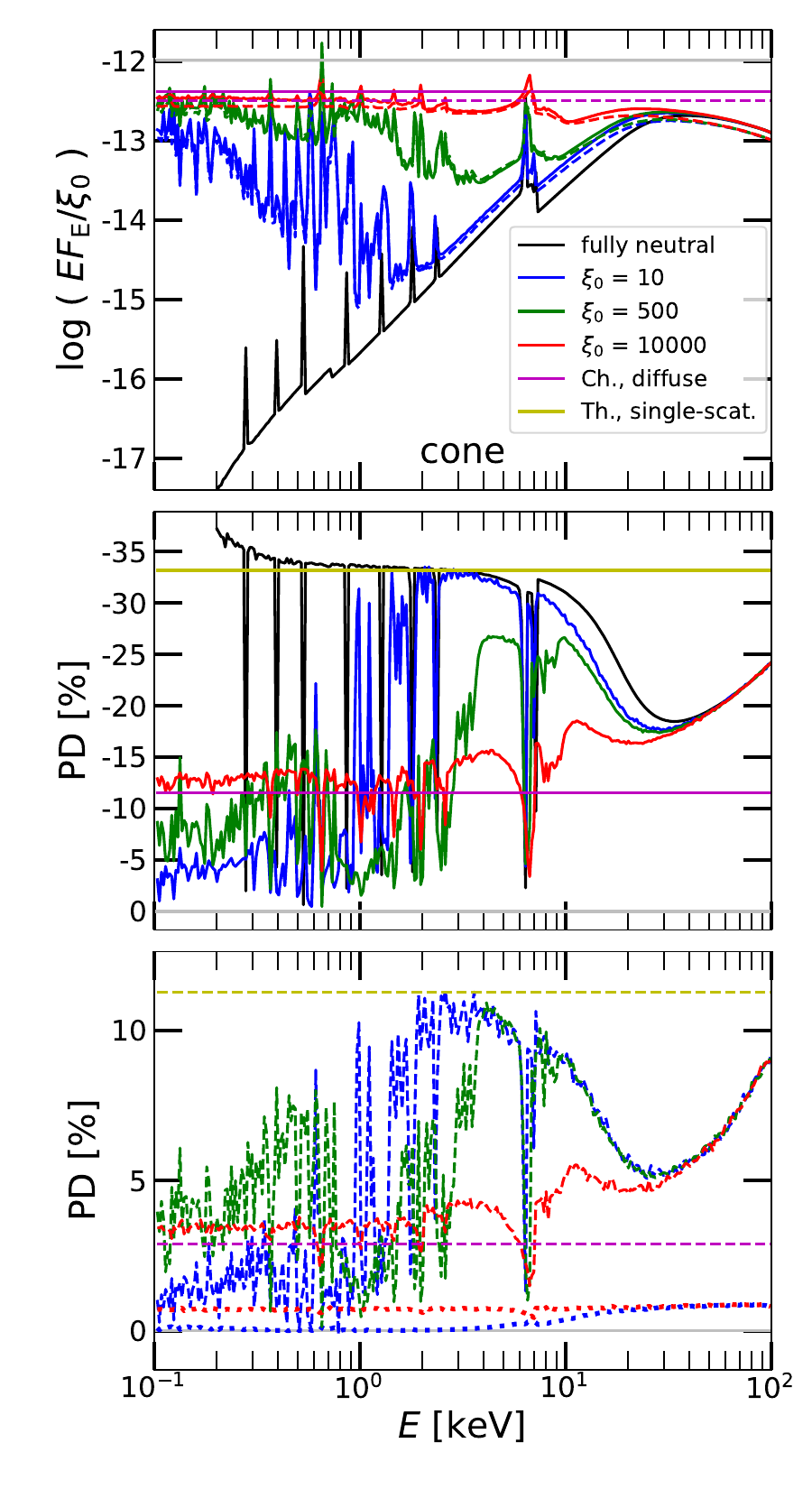}
%
\caption{\footnotesize The spectra (top panel) and the corresponding PD (middle and bottom panels) versus X-ray energy for a compact unpolarized isotropic power-law source (with $\mathrm{\Gamma} = 2$) illuminating an equatorial double-cone obscurer, viewed at 51$^\circ$ inclination from a 1 kpc distance. Solid lines show reflected-only emission from a cone with $\mathrm{\Theta} = 40^\circ$, $\rho = 1.8\,\rho_\mathrm{in}$ and $\rho_\mathrm{in} = 10^9\,\mathrm{cm}$, typically inducing a PA perpendicular to the projected axis of symmetry (negative PD). Dashed lines show reflected-only emission from a cone with $\mathrm{\Theta} = 70^\circ$, $\rho = 1.3\,\rho_\mathrm{in}$ and $\rho_\mathrm{in} = 10^9\,\mathrm{cm}$, typically inducing a PA parallel to the projected axis of symmetry (positive PD). The blue, green, and red curves correspond to different values of the ionization parameter $\xi_0$ in $[\textrm{erg} \,\,\textrm{cm} \,\,\textrm{s}^{-1}]$ at the inner edge of the double-cone. Special cases show the numerically calculated fully neutral reflection (black) and the classical analytical approximations for the fully neutral (yellow, in polarization only) and fully ionized (magenta) limits. The gray lines correspond to the primary radiation. The dotted lines in the bottom panel show the total (primary + reflected) polarization degree for two selected cases. Image adapted from \cite{Podgorny2025b}.}
\label{fig:torus_reflection}       
\end{figure}

Figure \ref{fig:torus_reflection} shows the re-processed spectra and PD versus energy for isotropic power-law illumination of a distant double-cone, with multiple ionization profiles shown in the color code to illustrate the dependence of the spectral and polarization results on the ionization state of the reflector. The spectral shapes again correspond to those of the underlying local reflection tables (see Figure \ref{fig:results_loc_reflection_xi}), and the fully ionized and fully neutral limiting cases obtained with numerical and analytical methods in identical geometry are shown for comparison. The PD values of the intermediate ionization cases obtained with the \texttt{STOKES\_CONE} model indicate that the combination of scattering, absorption, and spectral lines makes the dependence of PD on ionization non-trivial and highly energy-dependent, not monotonically varying between the fully neutral and fully ionized limits at all energies.

While the profile of PD with energy is determined by the underlying local reflection tables, its magnitude and sign are to a large extent determined by the global geometrical parameters $i$, $\mathrm{\Theta}$, and $\rho/\rho_\mathrm{in}$. The middle and bottom panels of Figure \ref{fig:torus_reflection} show two scenarios with smaller ($\mathrm{\Theta} = 40^\circ$) and larger ($\mathrm{\Theta} = 70^\circ$) opening angles, respectively, for a fixed observer at $i = 51^\circ$. In the first, highly obscured case, we obtain high PD fractions (tens of \%)  with a PA perpendicular to the projected axis of symmetry, because the overall symmetry is significantly reduced and the dominant scattering plane is meridional, with the prevailing PA set by the Thomson scattering law. Such polarization properties have been detected by {\it IXPE} in XRB and AGN sources thought to be engulfed in elevated equatorial obscurers, in particular Cygnus X$-$3 \cite{Veledina2024}, the Circinus Galaxy \cite{Ursini2023b}, and NGC 1068 \cite{Marin2024c}; although additional non-negligible contribution from reflection off polar matter to the observed polarization is anticipated \cite{Podgorny2024b, Veledina2024b}. The large opening angle of the obscurer on the other hand creates a situation that resembles the centrally illuminated disc reflection geometry and produces a reflected PA parallel to the projected symmetry axis, as the dominant scattering plane is equatorial. The PD values theoretically obtainable from pure reflection on a 3D toroidal geometry vary continuously between extreme negative values (close to 100\%) and intermediate positive ones (still tens of \%). The precise boundary at which the reflection-induced PD switches sign is given by a non-trivial contour in the $i$, $\mathrm{\Theta}$, $\rho/\rho_\mathrm{in}$ parameter space. 

\begin{figure}[t]
\sidecaption
\hspace*{5mm}\includegraphics[width=0.92\textwidth]{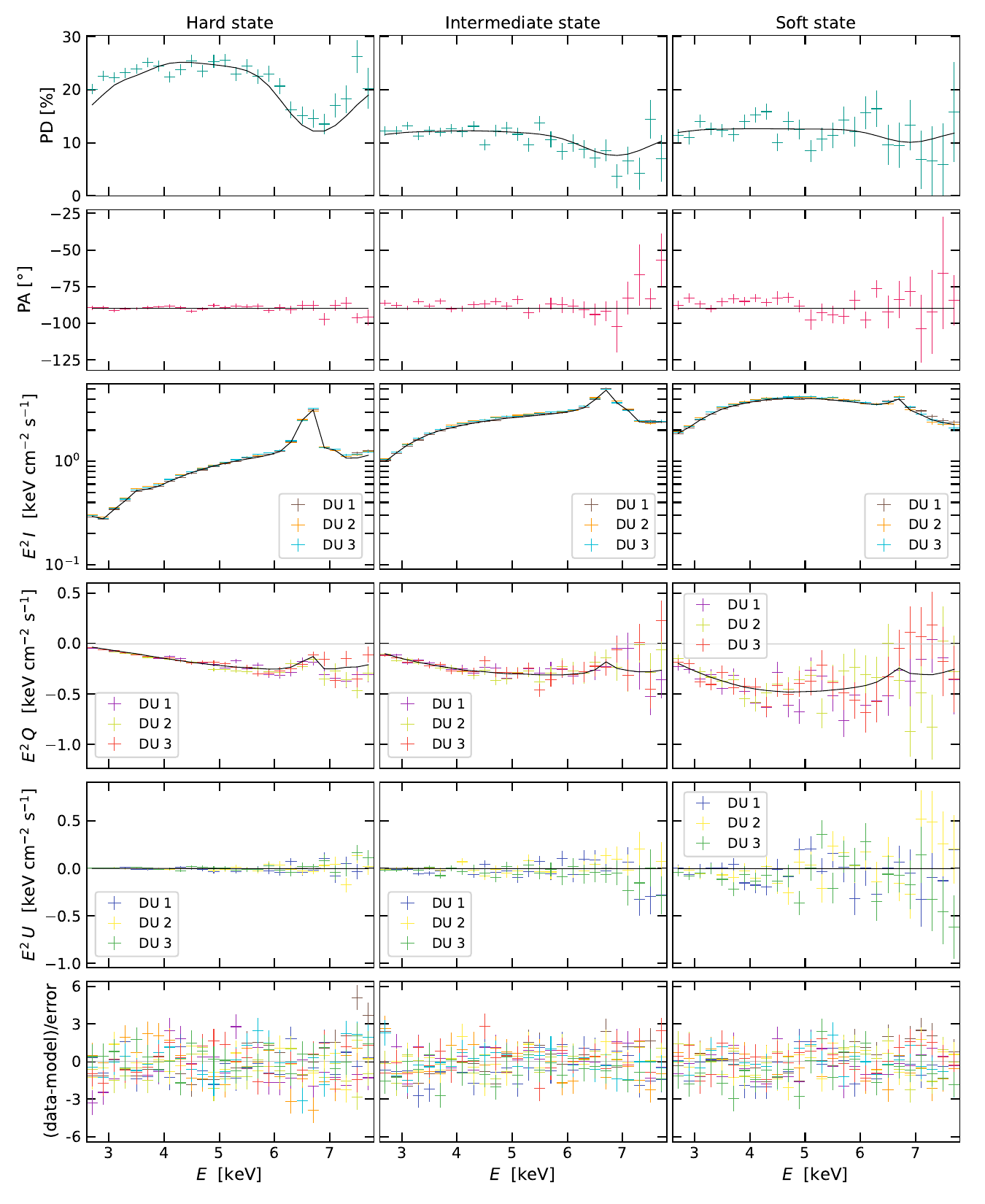}
%
\caption{\footnotesize Example of a fit in \texttt{XSPEC} to the 2022 {\it IXPE} data of Cygnus X$-$3 in the hard state (left) and intermediate state (middle) \cite{Veledina2024}, using the \texttt{STOKES\_BOWL} model; and of a similar fit to the 2024 {\it IXPE} data of its soft state (right) \cite{Veledina2024b}, using the \texttt{STOKESBB\_BOWL} model. The best-fit model is shown as the black solid line, and the data points are shown in colors. For clarity, the first two rows show the PD and PA with all DUs combined. The remaining rows show the Stokes $I$, $Q$, and $U$ spectra, multiplied by $E^2$, followed by the fit residuals, with all DUs shown separately. The ionization parameter $\xi_0$ increases from $\sim 100\,\,\textrm{erg} \,\,\textrm{cm}\,\, \textrm{s}^{-1}$ by several hundreds from the hard to the intermediate and soft states, while the accretion funnel geometry stays narrow, with the best-fit parameters favoring a shrinking cavity from the hard towards the soft state. Note that in this figure, $I$, $Q$, and $U$ 
denote Stokes parameters of the photon number density flux rather than the usual definition in terms of energy flux.\\[-8mm]}
\label{fig:Cyg_X-3_fit}       
\end{figure}
An application of the \texttt{STOKES\_TORUS} model to a composite {\it IXPE} spectro-polarimetric fit of an AGN is shown in Section \ref{polarization_KY_models}, in combination with the \texttt{KY} codes for the inner accretion region. Here, we provide a simpler example of a single-component fit in \texttt{XSPEC} using the \texttt{STOKES\_BOWL} and \texttt{STOKESBB\_BOWL} models for the first two {\it IXPE} observations of the Galactic BH XRB Cygnus X$-$3, obtained in the hard and intermediate states in 2022 \cite{Veledina2024} and in the soft state in 2024 \cite{Veledina2024b}. We fit the Stokes $I$, $Q$, and $U$ data from {\it IXPE} in the 2.5--8 keV band with the model combination \texttt{CONSTANT * TBABS * STOKES\_BOWL} or \texttt{CONSTANT * TBABS * STOKESBB\_BOWL}, in \texttt{XSPEC} notation. The \texttt{CONSTANT} accounts for differences in the absolute flux calibration among the three {\it IXPE} detector units (DUs), and \texttt{TBABS} \cite{Wilms2000} accounts for line-of-sight absorption, since the source is located in the Galactic plane at a distance of $\sim 10$ kpc. Cygnus X$-$3 is a Galactic ultra-luminous X-ray source with a narrow accretion funnel viewed at an inclination of $\sim 30^\circ$ \cite{Veledina2024}; therefore, only a single reflection component can crudely describe the X-ray data. With this model, we can test whether the observed drop in the 2--8 keV PD from $\sim 21\%$ to $\sim 10\%$ between the first two observations (with the PA remaining unchanged, perpendicular to the projected jet direction, and constant with energy) can be attributed to a change in the ionization of the inner funnel walls, while preserving their full opacity, as suggested by the spectra and the energy dependence of the PD. Figure \ref{fig:Cyg_X-3_fit} shows that such an interpretation is viable, although a detailed multi-instrument spectro-polarimetric analysis remains to be carried out. The best-fit parameter values and other fit details are provided in Table \ref{best-fit_Cyg_X-3} in the \nameref{technical_docs}. 

\bigskip
The \texttt{STOKES} tables described in this section, together with the non-relativistic \texttt{XSPEC} models that make use of them, are in most cases publicly available; links to the relevant repositories and documentation can be found in Tables \ref{tab:models1}--\ref{tab:models2} in the \nameref{technical_docs} section.

\section{KY codes}
\label{sec:2}

The \texttt{KY} codes\footnote{The package is named after Vladimír Karas and Tahir Yaqoob, who originally introduced them, with the subsequent development led predominantly by Michal Dovčiak over the past two decades.}
are \texttt{XSPEC}-compatible, fully relativistic spectro-polarimetric models for the inner accretion regions of XRBs and AGNs.  Individual models can be used either additively or multiplicatively within \texttt{XSPEC}, allowing combination with other models representing different physical components of the studied system. Each model can also be run outside \texttt{XSPEC} to produce raw unfolded model curves. For the most detailed treatment of local re-processing in disc atmospheres, the \texttt{KY} codes make use of the \texttt{STOKES} tables introduced in Section \ref{sec:1}. Because the implementation of relativistic ray-tracing and spectro-polarimetric imaging is to a large extent not specific to the X-ray band, some of the \texttt{KY} models extend their applicability to the UV and optical bands, for instance for broadband SED computations \cite{Dovciak2022} and reverberation mapping studies of AGN accretion discs \cite{Kammoun2023}.

Because all model variants in the \texttt{KY} package are based on the same relativistic ray-tracing, we first introduce its assumptions and physical set-up. We will not elaborate on the full mathematical apparatus, which can be found in the associated literature and code repositories. We then introduce each sub-model in order of increasing physical complexity, starting with the spectral and timing models and concluding with the polarimetric models that make use of the \texttt{STOKES} tables. Relevant examples of {\it IXPE} data fitting are shown for selected models.

\subsection{Relativistic ray-tracing}
\label{sec:ray-tracing}

The \texttt{KY} codes make use of pre-computed tables of numerically solved geodesic and geodesic deviation equations for photons in Kerr space-time parametrized by the dimensionless BH spin, $a \in [-1;1]$. The BH spin is defined as positive when the BH corotates with the accretion disc and negative when it counter-rotates. The disc is assumed to rotate counter-clockwise as seen by the distant observer. While the sense of the system rotation leaves no imprint on the observed spectrum, it does influence the predicted polarization angle. In the spectro-polarimetric models, the effective rotation direction can therefore be reversed by adopting a negative inclination, $i \in [-90^\circ, 0^\circ]$, which is equivalent to viewing the system from below and thereby flipping the apparent sense of rotation. Negative spin values are not currently implemented in the spectro-polarimetric models. 

\begin{figure}[t]
 \sidecaption
\makebox[0pt][l]{\raisebox{75pt}[0pt][0pt]{\hspace*{2pt}\footnotesize \,\,\,\,\,\,\,\bf{(a)}}}%
\includegraphics[scale=.152]{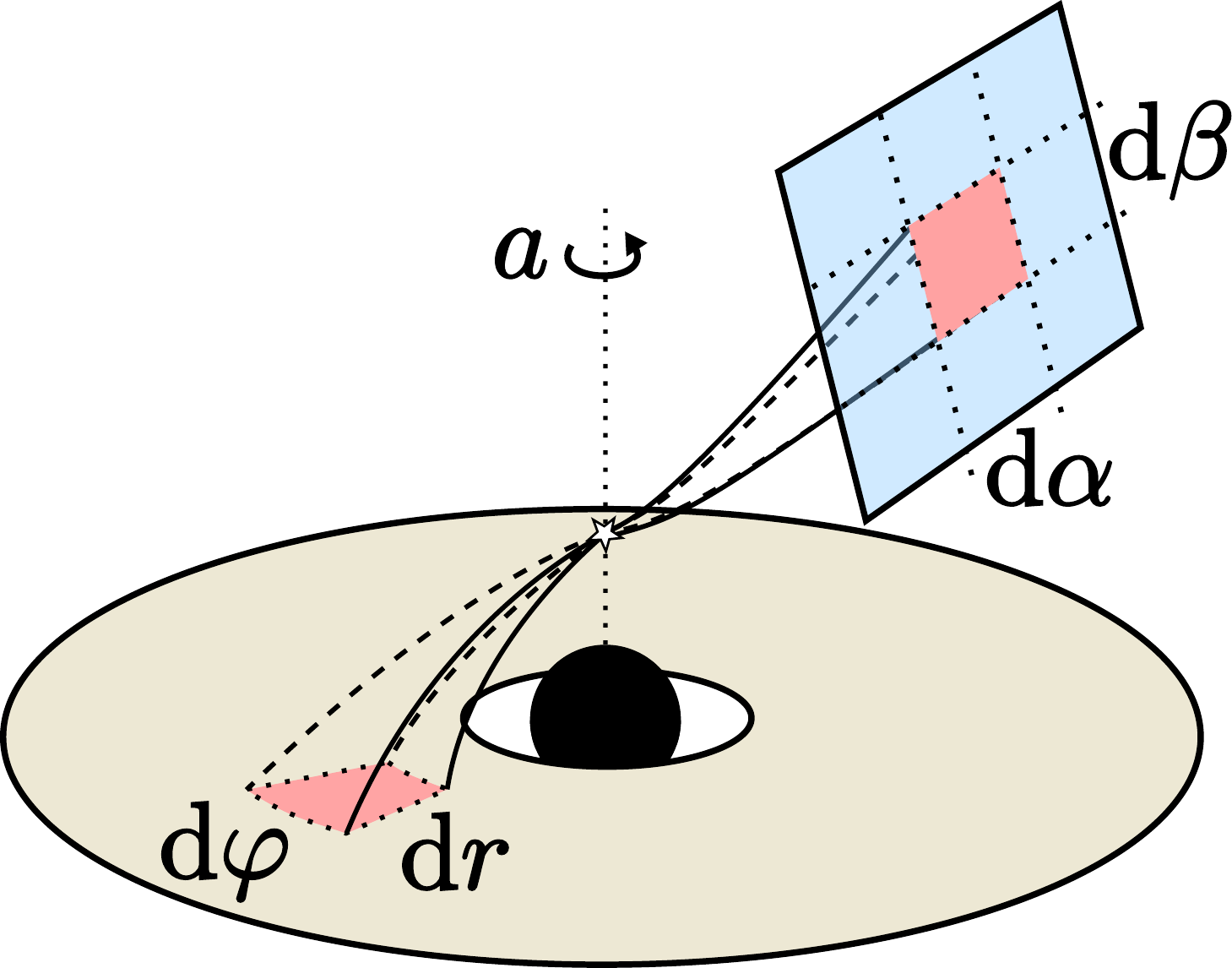}%
\hspace{0.5mm}
\makebox[0pt][l]{\raisebox{75pt}[0pt][0pt]{\hspace*{2pt}\footnotesize \,\,\,\,\,\,\,\bf{(b)}}}%
\includegraphics[scale=.152]{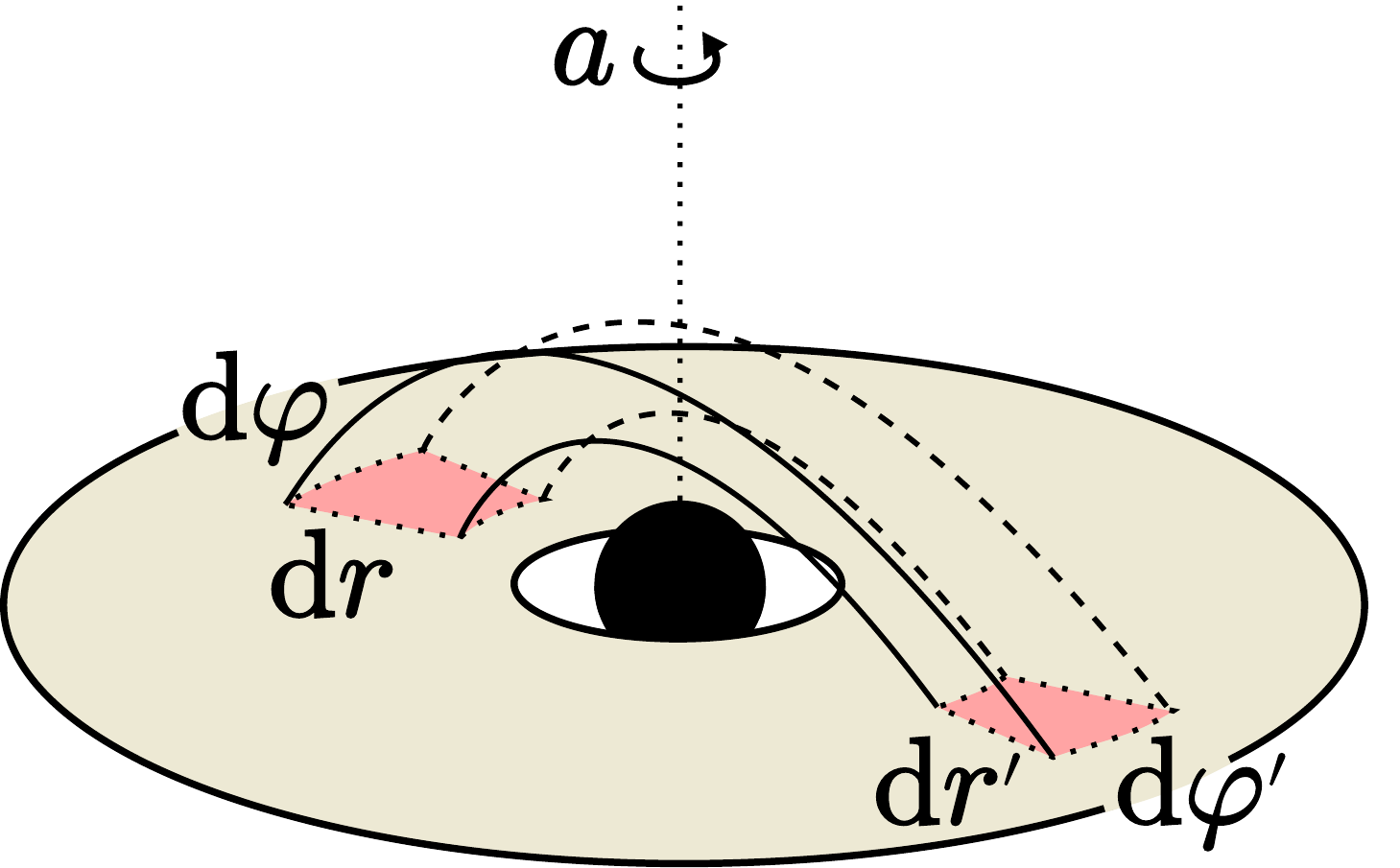}%
\hspace{1mm}
\makebox[0pt][l]{\raisebox{75pt}[0pt][0pt]{\hspace*{2pt}\footnotesize \,\,\,\,\,\,\,\bf{(c)}}}%
\includegraphics[scale=.152]{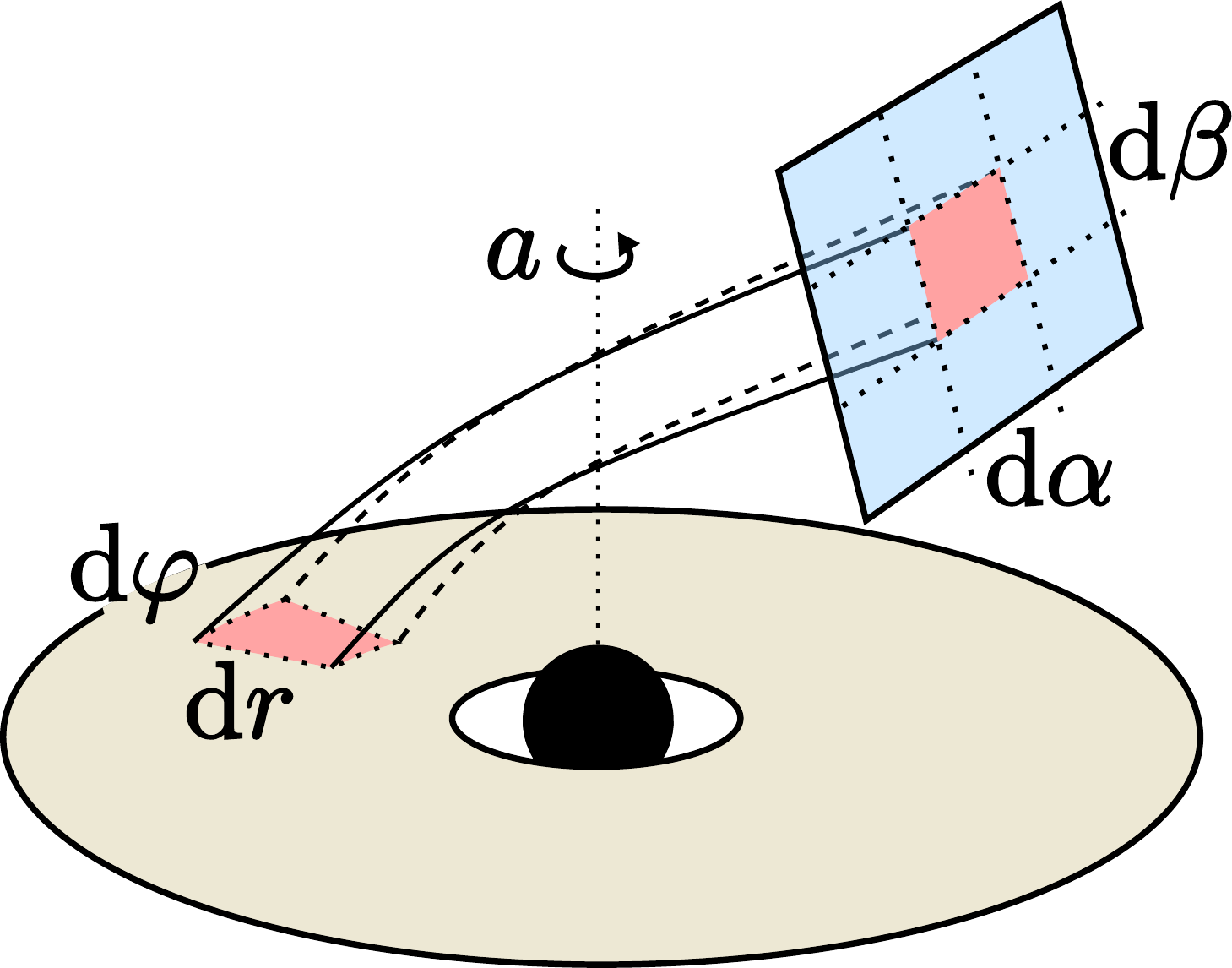}
%
\caption{\footnotesize Different cases of GR ray-tracing used in the \texttt{KY} codes: (a) geodesics between an on-axis point source (the lamp-post corona) and the equatorial plane (the accretion disc) or a distant observer at arbitrary inclination; (b) between one location in the equatorial plane and another (for the returning radiation from the disc or slab corona); and (c) between the equatorial plane (the disc or slab corona) and a distant observer at arbitrary inclination. Image courtesy of Matěj Ptáček.}
\label{fig:transfer_sketch}       
\end{figure}

The disc-corona system is treated as non-gravitating, i.e. its mass-energy contribution to the space-time curvature is neglected. The radiative transfer in matter is calculated in the local rest frame, either using (semi-)analytical prescriptions or by interpolating within the \texttt{STOKES} tables. This local treatment is separated from the global radiative transfer in vacuum between the axis, the disc, and the observer. Several types of global null geodesics are considered, as illustrated in Figure \ref{fig:transfer_sketch}: (a) between the BH axis and the equatorial plane or the distant observer (for the lamp-post corona illuminating the disc or emitting directly towards the observer, respectively), (b) between one location in the equatorial plane and another (for returning radiation, i.e. photons emitted at one disc location that are gravitationally bent back to another disc location), (c) between the equatorial plane and the distant observer (for observed disc thermal emission or disc reflection and, in some models, direct emission from an extended corona situated just above the disc).

The disc is considered optically thick and geometrically thin, strictly in the equatorial plane, and rotates with a Keplerian velocity profile. Unless stated otherwise, the Novikov-Thorne radial profile \cite{Shakura1973, Novikov1973} is assumed for thermal disc emission in the soft state of XRBs. The disc region below the innermost stable circular orbit (ISCO) may optionally be set to emit radiation, in which case the material is assumed to be in free fall with the energy and angular momentum of a particle orbiting at the ISCO. The coronal geometry is approximated in two extreme configurations. In the lamp-post geometry, the corona is represented as a static point-like isotropic source located on the BH rotation axis at height, $h$, above the disc. In the extended slab geometry, the corona is represented as a plane-parallel layer situated just above the accretion disc in the equatorial plane, with a radial power-law emissivity profile. The distant observer is located at spatial infinity at an arbitrary inclination, $i$, which coincides with the Boyer-Lindquist polar coordinate, $\theta$. Photons reaching the equatorial plane outside the disc edges, i.e. below the inner or beyond the outer disc radius, are not traced further, meaning that only first-order images of both the disc and the corona are considered, while higher-order images arising from photons crossing the equatorial plane multiple times are neglected. 

Time-variable properties of the system can be traced through the Boyer-Lindquist time coordinate $t$. The classical study \cite{Cunningham1973} introduced the impact parameters $\alpha$ and $\beta$ on the sky of the distant observer, which can be mapped to the equatorial plane covered with Boyer-Lindquist radial and azimuthal coordinates, $r$ and $\varphi$, respectively. Further, the \texttt{KY} codes allow the emission to be restricted to a particular disc segment in $r$ and $\varphi$, useful for investigating spot or arc emission, comparing emission from different disc regions or disc tomography studies. They also allow simulation of partial covering of the accreting system by a circular obscuring cloud, defined by its position in the impact parameters $\alpha$ and $\beta$ and its radius \cite{Kammoun2018}.

Although it is more common to discuss properties of the Kerr space-time in the Boyer-Lindquist coordinates, the full GR ray-tracing in the \texttt{KY} routines is technically (pre-)computed in the numerically better suited Kerr ingoing coordinates \cite{Dovciak2004, Dovciak2004a}. Radiation properties at the receiving point are computed by transforming the local photon momentum and polarization vectors from the rest frame of the emission point to global coordinates, and subsequently to the rest frame of the receiving point. The disc-plane integration
follows the method of \cite{Cunningham1976}, who introduced the transfer function --- a Jacobian relating the equatorial plane to the coordinates on the observer's sky. The transfer function encodes all relevant GR effects, including gravitational lensing and frequency shifting, and thereby assigns appropriate weights to different parts of the equatorial plane. When integrating over an emitting or reflecting accretion disc, the local emission function is multiplied by the transfer function in the integrand, yielding the well-known relativistically distorted imaging and line-broadening effects \cite{Dovciak2004, Dovciak2004a}.

The observed Stokes parameters $Q$ and $U$ are treated in the same manner as Stokes $I$. We calculate them from the locally emitted Stokes parameters $I'$, $Q'$ and $U'$ across an equatorial area $S$ in Boyer-Lindquist coordinates ($\mathrm{d}S=r\,\mathrm{d}r\,\mathrm{d}\varphi$), using the following expressions:
\begin{equation}
\begin{aligned}\label{totalpol}
\begin{split}
I(E) &= N_0\int I'\!(E\!/\!g)\,G\,\textrm{d}S\,,\\
Q(E) &= N_0\int [Q'\!(E\!/\!g) \cos(2\chi) - U'\!(E\!/\!g) \sin(2\chi)]\,G\,\textrm{d}S\,,\\
U(E) &= N_0\int [Q'\!(E\!/\!g) \sin(2\chi) + U'\!(E\!/\!g) \cos(2\chi)]\,G\,\textrm{d}S\,,
\end{split}
\end{aligned}
\end{equation}
where $N_0$ is the normalization factor accounting for the distance to the source and unit conversions and $G\equiv\mu_\mathrm{e}\,g^2l$ is the transfer function, with $\mu_\mathrm{e}\equiv\cos(\delta_\mathrm{e})$ being the cosine of the local emission angle, $l$ the lensing factor, defined as the ratio of the cross-section of the photon flux tube at infinity to that at the emitting frame, and $g$ the ratio of the observed to emitted photon energy, encoding both gravitational redshift and Doppler shifts. Here $\chi$ denotes the rotation of the PA between the local rest frame of the emitter and the distant observer's frame, which includes contributions from the parallel transport of the polarization vector in curved space-time, the transformation between locally defined frames, and the aberration of photon directions due to the relative motion of the emitter and receiver. The expressions for $Q$ and $U$ differ from that for $I$ only through this rotation angle $\chi$. When multiple geodesics contribute to the observed flux at the receiving point, each carries a different GR-induced PA rotation. Their superposition can therefore significantly modify the observed PD --- typically diluting it --- compared to the flat space-time expectation. To calculate $\chi$, methods from \cite{Connors1977, Connors1980} are used. The Walker-Penrose theorem \cite{Walker1970} allows for a definition of a complex constant, $\kappa$, at the emission point as well as at the receiving point. Expressing the parallel-transported polarization vector through $\kappa$ significantly simplifies the computation. Since $\kappa$ is conserved along null geodesics, it needs only to be evaluated analytically at the boundary locations relevant to the \texttt{KY} framework: the BH axis, the equatorial plane, and spatial infinity.

\subsection{Relativistic spectral models for XSPEC}

We introduce the \texttt{KY} sub-models in order of increasing complexity, beginning with those designed primarily for spectral fitting, followed by timing models and spectro-polarimetric models. This ordering reflects a practical consideration: the timing models are slower because the time coordinate introduces an additional computational dimension, making them unnecessarily costly for plain spectral analysis. The spectro-polarimetric models, while physically more advanced, operate on pre-computed tables with a lower energy resolution than their purely spectral counterparts, which is sufficient for the currently operating X-ray polarimeters such as {\it IXPE} that do not provide high spectral resolution, and results in faster computation. Because all \texttt{KY} models share the same underlying physical assumptions and set-up, a spectro-polarimetric fit performed with a polarimetric \texttt{KY} model can be directly complemented by substituting it in the same \texttt{XSPEC} set-up with a dedicated spectral \texttt{KY} model, when the higher spectral resolution demanded by simultaneously acquired data from, e.g., NICER or {\it NuSTAR}, is required (see, e.g., \cite{Kammouninprep}).

\subsubsection*{The relativistic line models}

Four \texttt{KY} models are suitable for the analysis of relativistically broadened spectral line profiles from BH accretion discs. They were motivated by earlier models of relativistic fluorescent line profiles \cite{Fabian1989, Laor1991}, which, unlike \texttt{KY}, treated the black hole spin as a fixed parameter. Historically, these were the first \texttt{KY} codes developed, introduced in \cite{Dovciak2004, Dovciak2004a, Dovciak2014}. \texttt{KYNRLINE} predicts the relativistic line profile of a reflecting disc with a broken power-law radial emissivity (using the photon paths depicted in Figure \ref{fig:transfer_sketch}c), allowing for several limb darkening and limb brightening laws for emission directionality. \texttt{KYNRLPLI} predicts the relativistic line profile from a reflecting disc in the lamp-post geometry (using the photon paths depicted in Figures \ref{fig:transfer_sketch}a and \ref{fig:transfer_sketch}c), where the fluorescent iron line is computed with the MC code \texttt{NOAR} \cite{Dumont2000} by subtracting the reflected continuum, giving slight limb brightening emission directionality that depends on the primary source power-law photon index. In all \texttt{KY} line models, only a single reflection from the disc is assumed; secondary reflections due to disc self-irradiation are not computed.

A popular alternative approach is a convolution of the locally emitted narrow line profile with the transfer function \cite{Dovciak2004}, which allows GR effects to be applied to any \texttt{XSPEC} spectral model, including the continuum [e.g., \texttt{KYNCONV(POWERLAW)}]. \texttt{KYNCONV} and \texttt{KYNCLP} are the convolution-model alternatives to \texttt{KYNRLINE} and \texttt{KYNRLPLI}, respectively. This approach, however, assumes the same local spectral shape across the entire disc, with only a radial and emission directionality dependence of the emissivity, thus neglecting any dependence on physical parameters such as the ionization parameter. It also significantly increases the computational cost compared to the direct line models \cite{Dovciak2004a}.

All four models include polarimetric predictions, introduced as a toy model to explore GR effects on polarization prior to the advent of X-ray polarimeters such as {\it IXPE}. They rely on the non-physical assumption that the local emission is 100\% linearly polarized perpendicular to the disc plane. Since fluorescent lines are intrinsically unpolarized, a more practical approach in \texttt{XSPEC} is to use the \texttt{POLCONST} model with zero polarization degree directly on the line component which will result in relativistically broadened dips in energy dependence of PD. We therefore do not recommend the built-in polarization of these models for quantitative spectro-polarimetric analysis.

\subsubsection*{Models for the disc reflection}
\label{sec:spectral}

\texttt{KYNREFIONX} and \texttt{KYNXILLVER} predict the full X-ray spectrum, including both the primary coronal emission and disc reflection with spectral lines, in the lamp-post corona geometry (using the photon paths depicted in Figures \ref{fig:transfer_sketch}a and \ref{fig:transfer_sketch}c) \cite{Dovciak2004, Dovciak2004a}. They assume partially ionized disc re-processing according to the rest-frame \texttt{REFLIONX} \cite{Ross1993, Ross1999, Ross2005} and \texttt{XILLVER} \cite{Garcia2010, Garcia2013, Garcia2014, Garcia2016} reflection tables, respectively, with several limb darkening and limb brightening laws for emission directionality implemented, if not directly predicted by the reflection tables. In the output, the primary coronal emission and disc reflection components may be provided separately or combined, which is useful for inspecting their relative contributions. Note that the local reflection tables are computed for a cut-off power-law incident spectrum, which should be kept in mind if a different prescription for the primary radiation is adopted. More recently, \texttt{KYNREFIONX} was extended to also allow a single power-law radial emissivity profile as an alternative to the lamp-post geometry, using the high-density \texttt{REFLIONX\_HD} tables \cite{Tomsick2018} for local disc re-processing \cite{Datta2024}.

In the extended slab corona geometry co-planar with the accretion disc (using the photon paths depicted in Figure \ref{fig:transfer_sketch}c), the \texttt{KYNHREFL} model provides only the reflected continuum, without the primary coronal emission or spectral lines \cite{Dovciak2004, Dovciak2004a}. It assumes a broken power-law emissivity profile and uses the multiplicative \texttt{XSPEC} model \texttt{HREFL} in the \texttt{HREFL*POWERLAW} syntax for local reflection of the incident power-law continuum, assuming locally isotropic disc illumination. \texttt{KYNHREFL} also provides polarimetric predictions, but similarly to the line models described above, it relies on the non-physical assumption of 100\% linear polarization perpendicular to the disc plane. We therefore do not recommend it for quantitative spectro-polarimetric analysis but recommend to use the \texttt{KYNLPCR} and \texttt{KYNSTOKES} models described in Section~\ref{polarization_KY_models} that include physically motivated polarization modeling instead.

In all disc reflection models, only a single reflection from the disc is assumed; secondary reflections due to disc self-irradiation are not computed, unlike the most recent version of the widely used \texttt{RELXILL} package \cite{Garcia2014, Dauser2014, Dauser2022}.

\subsubsection*{Disc-corona interaction models}

The interaction between the accretion disc and the corona can be more self-consistently studied with the \texttt{KYNSED} model, first introduced in \cite{Dovciak2022}. It couples the thermal disc emission with the Comptonized coronal emission through iterative computations of energy exchange between the two. It assumes the corona to be in the lamp-post geometry, hence the geodesics depicted in Figures \ref{fig:transfer_sketch}a and \ref{fig:transfer_sketch}c are used. The model does not account for returning radiation. In each iteration, radiation is propagated both from the lamp-post to the disc and from the disc to the corona. The disc is fixed to span between the ISCO and a user-defined outer radius, $r_\mathrm{out}$.

The output spectrum includes the thermal disc emission, the
Comptonized X-ray emission, as well as the X-ray reflection. The \texttt{XILLVER} tables  \cite{Garcia2014, Garcia2016} are used for the local disc reflection. Part of the illuminating flux that is not reflected is absorbed and thermalized; thereby, it increases the temperature of the disc. The disc consists of two regions divided by a transition radius, $r_\mathrm{t}$. Above $r_\mathrm{t}$, all energy produced by accretion is released in the form of thermal radiation (i.e. similar to the Novikov-Thorne disc but also including the thermalized energy due to illumination). Below $r_\mathrm{t}$, all energy released by accretion is channeled to the corona and fully used to heat the electrons in the corona that Comptonize the seed photons from the accretion disc. The thermal radiation of the disc below $r_\mathrm{t}$ is only due to the thermalization of the absorbed part of the X-ray illumination.

Although the \texttt{KYNSED} code itself does not compute Comptonization inside the corona, it modifies its total luminosity and low-energy cut-off in each iteration until convergence in both of these internal parameters is reached. The low-energy cut-off in the X-ray spectrum is computed by integrating the thermal disc photons that arrive at the corona to give the average seed photon energy. The total X-ray luminosity of the corona is given by the energy of incoming photons plus the energy dissipated by accretion of the Keplerian disc below the transition radius and above the ISCO. The number of scattered photons is conserved, and thus the size of the corona is also estimated. The coronal optical depth is estimated from the power-law index of the primary X-ray radiation. Unlike \texttt{KYNREFIONX} and \texttt{KYNXILLVER}, which are restricted to X-ray energies above 0.1~keV, \texttt{KYNSED} also provides output below this threshold, covering the UV and optical bands where the thermal disc emission — the so-called UV bump — dominates the AGN spectral energy distribution. This makes it well suited for broadband fitting of X-ray, UV, and optical data of AGN \cite{Kammoun2024}.

\subsection{Relativistic timing models}

\subsubsection*{X-ray reverberation models} 

Several \texttt{KY} models are designed to predict the timing properties of AGN and XRBs in the lamp-post regime. The outputs of the models (when used outside of \texttt{XSPEC}) are the time-dependent spectra of the disc response to an observed primary flash, the integrated spectrum and the light curve for a given energy range, the lag as a function of frequency between given energy bands, and the lag as a function of energy for different frequencies. The real and imaginary parts, amplitude, and phase of the Fourier transform of the response function are also provided, allowing for a more complete characterization of the variability properties of the system. These models share the same physical assumptions and ray-tracing framework as the spectral and spectro-polarimetric \texttt{KY} models, making them well suited for complementary timing analyses of sources also studied with the spectral or polarimetric \texttt{KY} codes.

The \texttt{KYNREVERB} model suite computes the time-dependent reflection spectra of the disc as a response to a flash of primary power-law radiation from the lamp-post location (using the photon paths depicted in Figures \ref{fig:transfer_sketch}a and \ref{fig:transfer_sketch}c). Its two model variants \texttt{KYNREFREV} and \texttt{KYNXILREV} are the reverberation counterparts of the spectral models \texttt{KYNREFIONX} and \texttt{KYNXILLVER}, using the \texttt{REFLIONX} \cite{Ross1993, Ross1999, Ross2005} and \texttt{XILLVER} \cite{Garcia2010, Garcia2013, Garcia2014, Garcia2016} tables for the local disc re-processing, respectively. Note that the extension to the high-density \texttt{REFLIONX\_HD} tables \cite{Tomsick2018, Datta2024} is not available in \texttt{KYNREFREV}. In each variant, the increase in the disc temperature due to partial thermalization of the illuminating flux is accounted for, but no iterative disc-corona interaction is performed. The models are based on the \cite{Dovciak2004, Dovciak2004a} computational scheme and have been discussed and used for AGN X-ray reverberation studies, e.g., in \cite{Caballero2018, Alston2020}.

The \texttt{KYNXILTR} model is based on the \texttt{KYNSED} model and has been introduced in \cite{Kammoun2023}. In the same iterative disc-corona scheme as \texttt{KYNSED}, it computes the response function of the accretion disc illuminated by the lamp-post corona. The \texttt{XILLVER} tables \cite{Garcia2014, Garcia2016} are used for rest-frame disc re-processing. This model can be used to simulate also the UV and optical disc response functions or to fit the observed UV and optical time lags as a function of wavelength, extending its direct application beyond the X-ray band \cite{Kammoun2023, Langis2024}.

\subsubsection*{Orbiting spot models}

The works \cite{Dovciak2004c, Dovciak2004a, Dovciak2008b, Dovciak2007} introduced the \texttt{KY} computational scheme for orbital analysis of the relativistic spectral features from X-ray-illuminated spots on the accretion disc surface, allowing calculation of time-dependent spectra and light curves of localized flares in the equatorial plane. The scheme was subsequently applied to model near-infrared flares from the Galactic centre supermassive black hole Sgr~A*, including polarimetric studies \cite{Meyer2006a, Meyer2006b, Eckart2008, Zamaninasab2010, Kunneriath2010}. However, the associated \texttt{KYNSPOT} model has not yet been translated from its original Fortran77 implementation to C and is therefore not currently available as part of the \texttt{KY} package.

\subsection{Relativistic spectro-polarimetric models for XSPEC}\label{polarization_KY_models}

This subsection introduces the \texttt{KY} models with polarization capabilities. Some of these models rely on simple (semi-)analytical approximations for the local polarimetric properties, while the more advanced ones use re-processing tables computed with the \texttt{STOKES} code, i.e.\ the \texttt{STOKES}, \texttt{STOKESBB} or \texttt{STOKESBBTRANS} table variants. While any of these models may also be used for plain X-ray spectral fitting in \texttt{XSPEC}, their energy resolution and physical complexity may be limited relative to the standards of contemporary cutting-edge X-ray spectrographs. On the other hand, the energy resolution for the Stokes parameters $I$, $Q$, and $U$ is far beyond what is needed for the currently operating and forthcoming X-ray polarimeters alone (see Part 4). 

The \texttt{KY} polarization models predict the spectrum and polarization properties of the inner accretion region as a function of the observer's viewing direction, making them well suited for use as input into parsec-scale Monte Carlo simulations with the \texttt{STOKES} code (or others). There, the inner region emission can be combined with distant re-processing components to predict spectro-polarimetric observables for the current and forthcoming X-ray polarimeters \cite{Marin2018c, Marin2018b, Podgorny2024b}.

\subsubsection*{Thermal disc emission and reflection of thermal returning radiation}

The \texttt{KYNBB} model suite, originating from the \texttt{KY} computational scheme \cite{Dovciak2004, Dovciak2004a}, focuses on modeling the spectral and polarization properties of the observed thermal multi-temperature blackbody radiation from the accretion discs of XRBs. For the local polarization of the direct emission, the suite supports both Chandrasekhar's analytical formulae for a semi-infinite electron-scattering atmosphere and tabulated transmission results for a finite optical depth atmosphere pre-computed with \texttt{STOKES} code. The first \texttt{XSPEC}-compatible model for the polarization of direct thermal radiation from a geometrically thin optically thick disc was presented in \cite{Dovciak2008}, extending the calculations of \cite{Connors1977, Connors1980} by including the spectral hardening factor and several values of the optical depth of the electron-scattering atmosphere (pre-computed in a tabular form with the \texttt{STOKES} code). Note that in the $\tau$-dependent \texttt{STOKESBBTRANS} (v1.0) tables used, the atmosphere is assumed to be a symmetric slab illuminated from the centre (as in \cite{SunyaevTitarchuk1985}), whereas the more physical geometry of a disc atmosphere illuminated from below was adopted in the subsequent works \cite{Taverna2021, Ratheesh2024, Marra2025} (\texttt{STOKESBBTRANS} v2.0 and higher). Since GR effects produce a characteristic energy dependence of both PD and PA, with the energy dependence of the PA depending quite strongly on the BH spin, the model can be used to fit the polarimetric data to estimate the BH spin value of XRBs in the soft state. The \texttt{KYNPHEBB} variant differs from \texttt{KYNBB} only in assuming a phenomenological radial power-law profile of the disc temperature. Both models compute only the direct radiation (see Figure \ref{fig:transfer_sketch}c).
\begin{figure}[b]
 \sidecaption
\includegraphics[scale=.49]{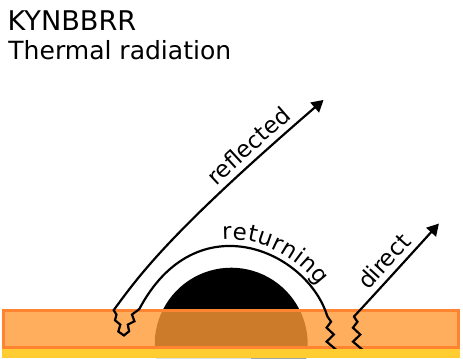}
\includegraphics[scale=.49]{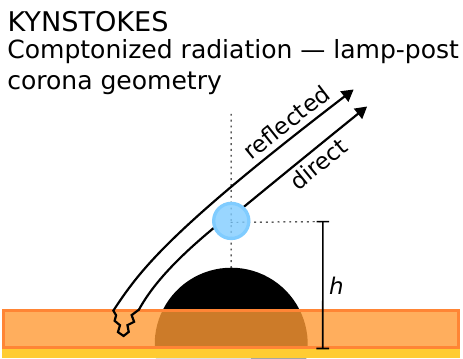}
\includegraphics[scale=.49]{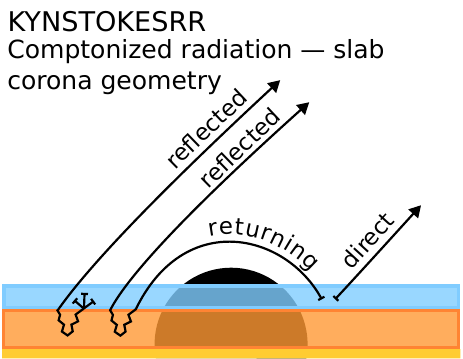}
%
\caption{\footnotesize Sketches of the different emission-reflection geometries used by the polarization \texttt{KY} codes. The \texttt{KYNBBRR} code (left panel) in its current version includes direct thermal radiation from the disc scattered through the disc atmosphere and (once) returning thermal radiation re-processed in the disc atmosphere. The \texttt{KYNSTOKES} code (middle panel) in its current version includes direct Comptonized radiation from the lamp-post corona and (once) re-processed radiation in the disc atmosphere. The \texttt{KYNSTOKESRR} code (right panel) in its current version includes direct Comptonized radiation from the slab corona, locally re-processed radiation in the disc atmosphere and (once) returning radiation re-processed in the disc atmosphere. Image courtesy of Matěj Ptáček.}
\label{fig:ky_geometries}       
\end{figure}

\begin{figure}[h]
 \sidecaption
\includegraphics[scale=.34]{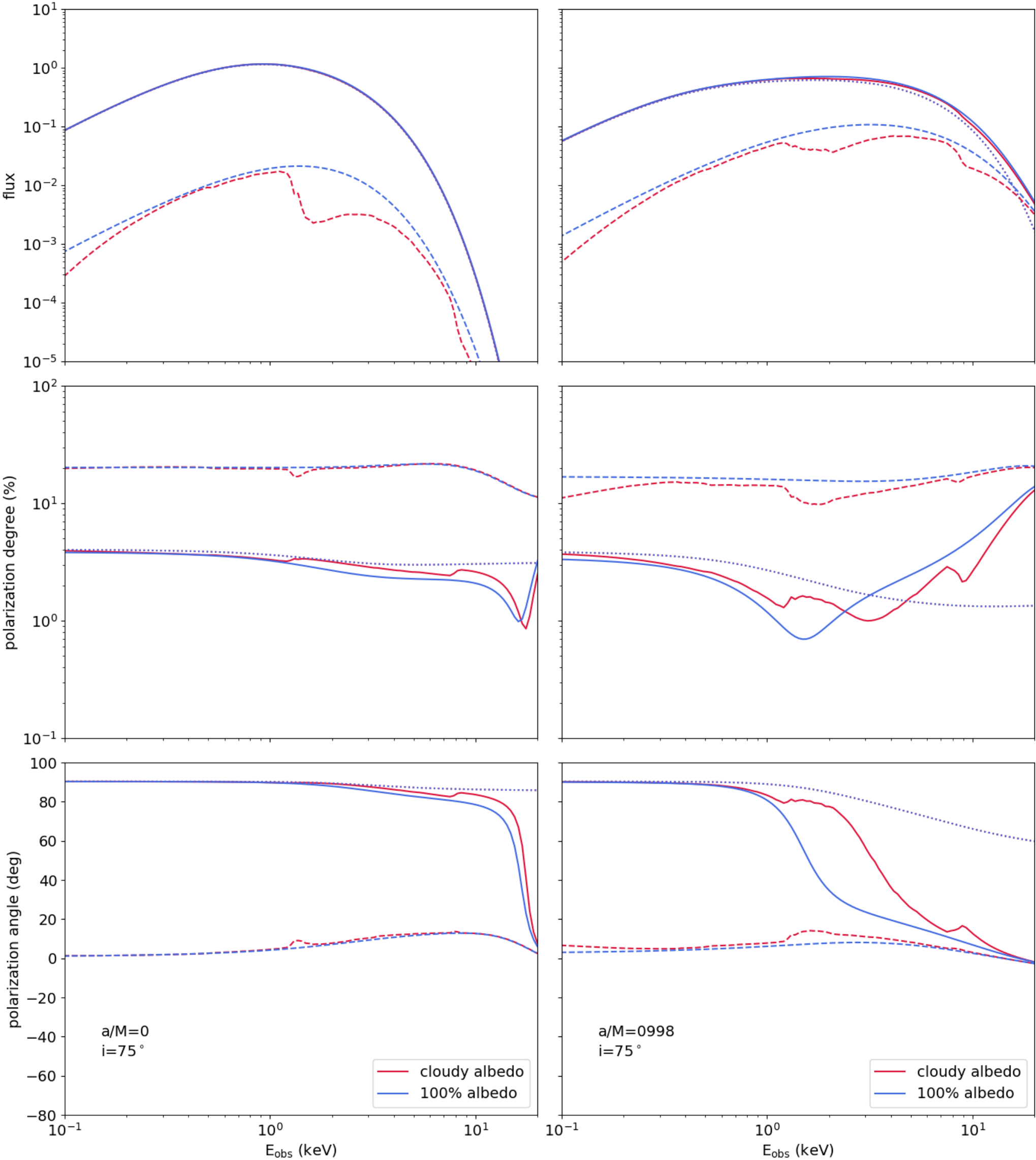}
%
\caption{\footnotesize The direct thermal emission from the disc re-processed in the electron-scattering semi-infinite disc atmosphere (dotted lines), the 
returning thermal radiation reflected from the disc atmosphere (dashed lines), and the total (direct + returning) radiation as seen by a distant observer at $75^\circ$ inclination, calculated with \texttt{KYNBBRR} in a variant adapted for tests of different albedo profiles. We provide cases with an energy-dependent albedo profile from \texttt{CLOUDY} (red lines) and an energy-independent 100\% albedo (blue lines). We show the observed spectra (top row), polarization degree (middle row), and the polarization angle (bottom row). We use BH mass $M = 10M_\odot$, hardening factor $f_\mathrm{col} = 1.8$, and atmospheric column density $N_\mathrm{H} = 10^{24}\, \mathrm{cm}^{-2}$ for all shown examples. The results are provided for BH spin $a=0$ with accretion rate $\dot{M} = 2.45 \times 10^{18} \,\mathrm{g/s}$ (left column) and $a = 0.998$ with $\dot{M} = 0.35 \times 10^{18} \,\mathrm{g/s}$ (right column). Image adapted from \cite{Taverna2020}.}
\label{fig:kynbb_RR}       
\end{figure}

The \texttt{KYNBBRR} model assumes the Novikov-Thorne temperature profile and, in addition to direct radiation, accounts for the contribution from returning radiation (i.e., ray-tracing along geodesics of all three types shown in Figure \ref{fig:ky_geometries}, left panel, is performed). The returning radiation is reflected from the disc only once — secondary reflections are not considered. The returning radiation was introduced in \cite{Taverna2020}, which provided cross-validation of the returning-radiation effects against the original results of \cite{Schnittman2009}, and evaluated the role of energy-dependent albedo profiles computed in CIE with \texttt{CLOUDY}, computed with a special publicly unavailable variant of the \texttt{KYNBBRR} code. Figure \ref{fig:kynbb_RR} shows the observed spectra and polarization for different prescriptions for the re-processing of the returning radiation and for different BH spins and accretion rates. The PA follows a characteristic swing in the X-ray band due to the competing almost orthogonally polarized direct and reflected returning radiation components, with both polarization directions being additionally rotated from their intrinsic values by GR effects along their respective geodesics. Note that while \cite{Taverna2020} studied energy-dependent albedo profiles, the 
main \texttt{KYNBBRR} model assumes a constant albedo for the reflection of returning radiation. Examples of spectro-polarimetric predictions of the \texttt{KYNBBRR} model for different BH spins, system inclinations, and accretion disc albedo values can be found in \cite{Mikusincova2023} while examples of spectro-polarimetric fits of the {\it IXPE} data with this model can be found in, e.g., \cite{Podgorny2023, Marra2023, Svoboda2024b, Qing-Chang2026}.

Two simpler variants of the model with returning radiation 
will soon become publicly available: the original \texttt{KYNBBRR} model, which assumes a full standard disc extending to the ISCO and uses only Chandrasekhar's formulae for the local polarization of the direct emission, and the more complete \texttt{KYNFBBRR}, which additionally supports the $\tau$-dependent \texttt{STOKESBBTRANS} tables with elastic scattering (v1.0) for the direct emission and includes further geometrical parameters. For the reflection of returning radiation, both variants offer three options. The first is a simple isotropic blackbody reflection with Chandrasekhar's formulae for the polarization, where the amount of reflection is controlled by the albedo parameter. The second uses the \texttt{XILLVERNS} tables \cite{Garcia2022} for the reflected spectral shape combined with Chandrasekhar's formulae for the polarization, and the third uses the \texttt{STOKESBB} reflection tables for the full Stokes parameter treatment. In the latter two cases, the amount of reflection and the shape of the reflected spectra are governed by the ionization parameter $\xi$, with a higher ionization leading to more reflection. In the \texttt{XILLVERNS} case, the spectral shape is further modified by the disc density and iron abundance. All three options depend on the local blackbody temperature of the returning radiation, which is shifted to higher values due to GR blueshift along the returning geodesic and may significantly exceed the local disc temperature at the point of reflection. The \texttt{XILLVERNS} option is computationally faster than the \texttt{STOKESBB} tables, although the two give slightly different predictions for the PD. In all three cases, it is also possible to compute only the returning radiation component, excluding the direct emission.

\begin{figure}[t]
\includegraphics[width=\textwidth]{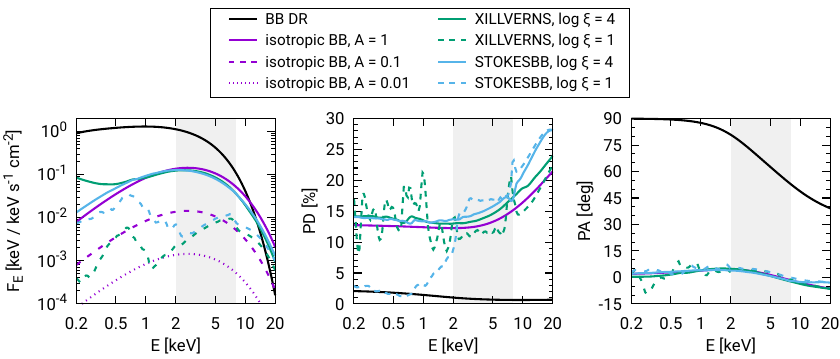}
%
\caption{\footnotesize \texttt{KYNBBRR} prediction for observed energy flux (left), polarization degree (middle) and polarization angle (right) for direct radiation (black, solid) and three assumptions for reflection of returning radiation: isotropic blackbody reflection with Chandrasekhar approximation for the polarization (purple), \texttt{XILLVERNS} reflection tables also with Chandrasekhar approximation for the polarization (green), and \texttt{STOKESBB} reflection tables (light blue) with self-consistent computation of polarization properties. While the blackbody reflection assumes albedo $A=1$ (solid), 0.1 (dashed) and 0.01 (dotted), the reflection tables assume the ionization parameter $\xi=10000$ (solid; highly ionized) and $10$ (dashed; almost neutral). Other parameters used: system inclination $i=60^\circ$, BH mass $M=10\,M_\odot$, BH spin $a=0.998$, accretion rate $\dot{M}=0.1\,\dot{M}_{\rm Edd}$, color correction factor $f_{\rm col}=1.7$, distance $D=10\,$kpc, and the \texttt{XILLVERNS} tables were used with solar abundances and disc density of $10^{15}$cm$^{-3}$. Energy band covered by the {\it IXPE} mission is denoted by gray area.}
\label{fig:kynbb_RR2}       
\end{figure}

\begin{figure}[b]
 \sidecaption
\hspace*{2em}\includegraphics[scale=.9]{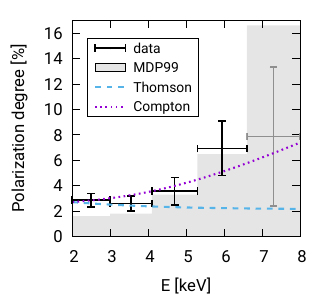}
\hspace*{2em}\includegraphics[scale=.9]{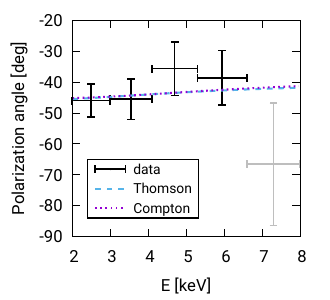}
%
\caption{\footnotesize Example of usage of the \texttt{KYNBBRR} model for spectro-polarimetric fitting of the LMC X$-$3 {\it IXPE} data in \texttt{XSPEC}. On the left we show the observed polarization degree (with gray bars corresponding to the $\mathrm{MDP}_{99\%}$ in each energy bin), and on the right the observed polarization angle. The source is in the soft state, with nearly pure thermal emission dominating the 2--8~keV band. 
The light-blue dashed lines are the best-fit model with Thomson electron-scattering atmosphere. 
The purple dotted lines are the same but with a highly-ionized Compton-scattering cold atmosphere of high optical depth described by the \texttt{STOKESBBTRANS} v3.0 tables pre-computed by the \texttt{TITAN} and \texttt{STOKES} codes, which provides a more favorable fit to the data in this case. In both cases, the reflection of returning radiation was not needed (zero albedo was assumed). 
Image adapted from \cite{Svoboda2024}.}
\label{fig:kynbb_direct_LMCX3}       
\end{figure}

An example of the three reflection prescriptions is shown in Figure~\ref{fig:kynbb_RR2}. The simple isotropic blackbody reflection provides a basic approximation without any spectral features, which may be sufficient for a highly ionized disc where the reflected spectrum is dominated by the continuum. The \texttt{XILLVERNS} tables provide a more realistic description of the reflected spectral shape, including disc re-processing features, and are computationally fast, making them particularly useful for spectral fitting of the reflection of returning radiation. However, the polarization properties in this case are computed with Chandrasekhar's approximation for elastic scattering applied to the full reflected spectrum without distinguishing between its components. This means that the increase in PD due to inelastic scattering and due to absorption is not captured, the fluorescent emission lines are assumed to be polarized as the reflecting continuum despite being intrinsically unpolarized, and the blackbody component arising from thermalization of the illuminating flux --- visible below $\sim$0.5~keV in Figure~\ref{fig:kynbb_RR2} --- is treated in polarization as scattered reflection. The \texttt{STOKESBB} blackbody reflection tables, despite lacking the thermalized soft component and Comptonization on hot electrons in the current version, provide the most physically complete treatment of the polarization properties, but are computationally expensive and are therefore best suited for checking and validating predictions made with the other two prescriptions. As shown in Figure~\ref{fig:kynbb_RR2}, the three prescriptions give very similar results for a highly ionized disc (high albedo or high $\xi$), with only minor differences in spectral shape due to Compton scattering on cold electrons. For lower ionization, the spectral shapes differ more significantly, particularly below $\sim$2~keV, which is at least partly attributable to differences in the atomic data used in the respective table computations — a known issue when comparing different reflection codes \cite{Podgorny2022,Podgorny2025}. The predicted PD also converges between the three prescriptions for a highly reflecting disc, while it may differ significantly otherwise. Notably, however, the predicted PA is similar across all three cases regardless of the reflection prescription, which suggests that the energy dependence of the PA is driven primarily by GR effects rather than by the details of the local reflection treatment.

\begin{figure}[t]
 \sidecaption
\includegraphics[scale=.39]{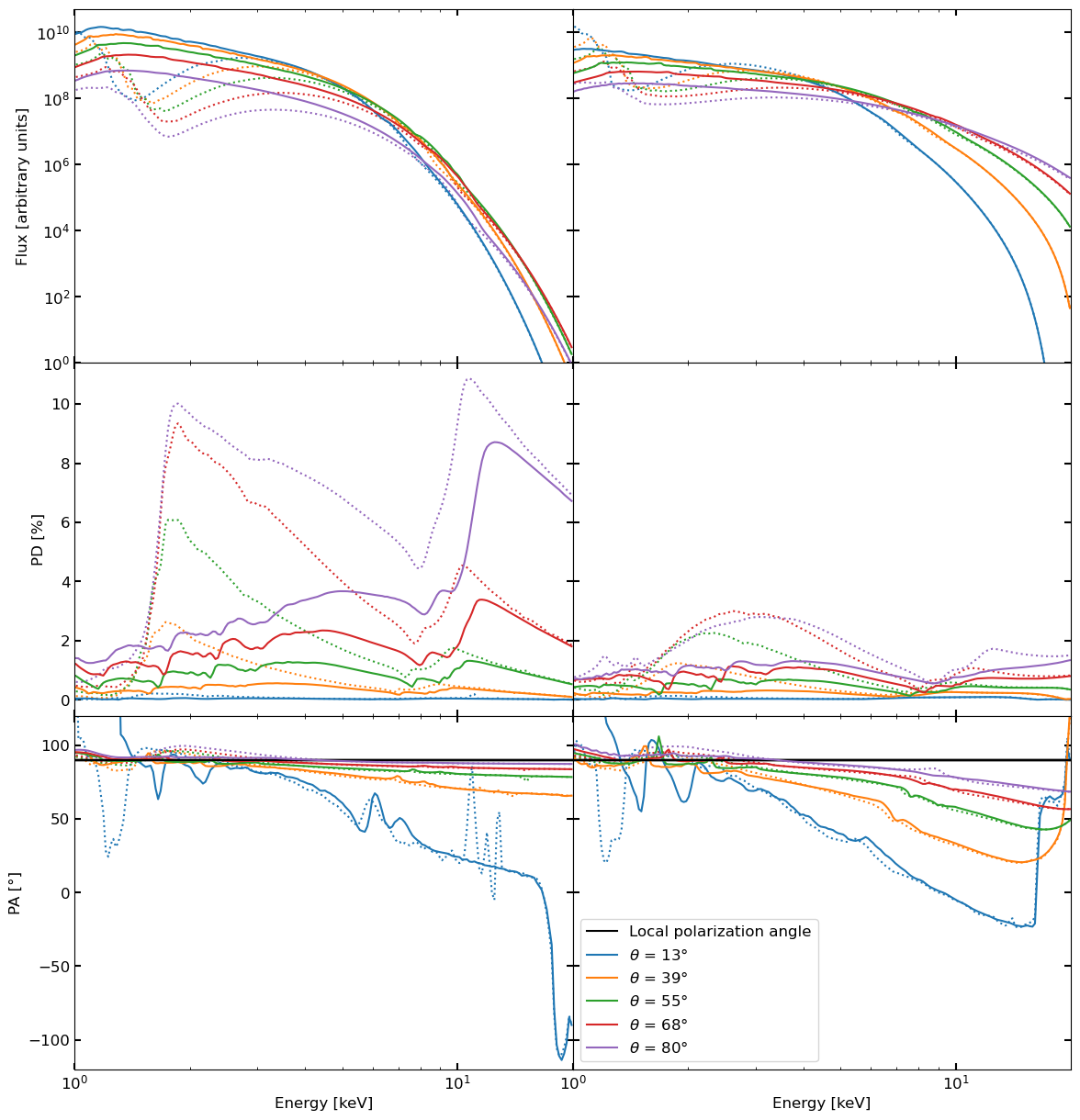}
%
\caption{\footnotesize The direct thermal emission from the accretion disc re-processed in the disc atmosphere assuming PIE (solid lines) and CIE (dotted lines), as predicted by the updated \texttt{KYNBBRR} variant with improved atmospheric treatment including absorption effects, for a distant observer. The inclination of the observer is given by the color code. We show the observed spectra (top row), polarization degree (middle row), and the polarization angle (bottom row). The black lines indicate the locally emergent polarization angle, which is perpendicular to the projected axis of symmetry. The results are provided for BH spin $a=0$ (left column) and $a = 0.998$ (right column). We use BH mass $M = 10\,M_\odot$, accretion rate $\dot{M} = 0.1\,\dot{M}_\mathrm{Edd}$, hardening factor $f_\mathrm{col} = 1.8$, and atmospheric column density $N_\mathrm{H} = 10^{24}\, \mathrm{cm}^{-2}$ for all shown examples. Image from \cite{Marra2025}.}
\label{fig:kynbb_direct}       
\end{figure}
While the above describes the different options for modeling the reflection of returning radiation, the treatment of absorption in the atmosphere for the direct emission has also been progressively refined in subsequent works. The main absorption features occurring in the CIE regime were studied in \cite{Taverna2021} using the \texttt{CLOUDY} and \texttt{STOKES} codes in a limited range of optical depths. The modeling in \cite{Ratheesh2024}, where the \texttt{TITAN} code was used in the PIE regime, showed consistency with the CIE approach of \cite{Taverna2021} in the overlapping regions of the parameter space. That work also extended the \texttt{STOKESBBTRANS} transmission tables to a wider range of $\tau$ for the nearly fully ionized scenario using the \texttt{TITAN} and \texttt{STOKES} codes, resulting in their v3.0 variant that includes inelastic scattering on cold electrons.

The model variant with the highly ionized \texttt{STOKESBBTRANS} v3.0 tables\footnote{This model variant is referred to as \texttt{KYNEBBRR} in \cite{Svoboda2024} but is not yet publicly available under that name and may be fused into the main \texttt{KYNBBRR} model in the future.} was successfully applied in the study of the LMC X$-$3 XRB in the soft state observed by {\it IXPE} \cite{Svoboda2024}, where it required a high atmospheric optical depth ($\tau\sim5$). This model provided a statistically more favorable fit than the original \texttt{KYNBBRR} model, as shown in Figure \ref{fig:kynbb_direct_LMCX3}. It better reproduces the increase of polarization degree with energy due to the increased role of down-scattering in the transmitting atmospheric layers with high $\tau$. The same local effect has been introduced in Figures \ref{fig:trans_BB} and \ref{fig:trans_tau}; GR effects largely preserve this local increase of PD with energy. Assuming the standard Novikov-Thorne disc, the same updated model also allowed successful polarization fits of the {\it IXPE} data of 4U1630$-$47 XRB in the soft state \cite{Ratheesh2024}, where, however, an outflowing atmosphere at relativistic speeds had to be additionally assumed to explain very high levels of polarization degree. We also remind that a self-consistent calculation of a locally thermally emitting atmosphere with a vertical distribution of electron temperatures, densities, and radiation sources remains to be done.

The absorption effects in local transmission were further evaluated for the direct thermal emission in \cite{Marra2025}, representing the most recent effort to model a more realistic disc atmosphere that is not assumed to be nearly fully ionized as in the \texttt{STOKESBBTRANS} v3.0 tables. Unlike the highly ionized case, where  only the PD is modified, a partially ionized atmosphere also changes the spectral shape of the transmitted flux due to absorption. Again, the local transmission tables were computed with the coupled \texttt{CLOUDY} and \texttt{STOKES} codes for both the PIE and CIE regimes. In Figure \ref{fig:kynbb_direct} we compare these two regimes for different system inclinations and BH spins from the perspective of a distant observer.

\begin{figure}[t]
 \sidecaption
\includegraphics[scale=1.]{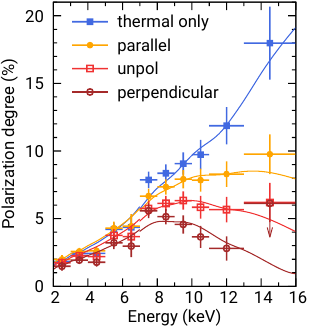}
%
\caption{\footnotesize Simulated 600\,ks observation of the polarization degree of 4U\,1957+115 in the soft state with the proposed {\it EXPO} mission, computed with the {\tt KYNBBRR} model. The blue curve shows the pure thermal component; the orange, red, and dark red curves show the effect of a weak Comptonized contamination that is polarized at 5\% parallel to the system axis, unpolarized, and polarized at 5\% perpendicular to the system axis, respectively.
}
\label{fig:EXPO_sim}       
\end{figure}
Beyond fitting current data, the {\tt KYNBBRR} model can equally be used to simulate observations with future X-ray polarimetric missions. As an example, Figure~\ref{fig:EXPO_sim} shows a simulated {\it EXPO} \cite{Soffitta2026} observation of the soft-state BH XRB 4U\,1957+115. The broad energy coverage of {\it EXPO}, extending to $\sim$35\,keV, is particularly valuable for soft-state sources: it simultaneously probes the high-energy rise of polarization degree expected from disc self-irradiation in the case of high BH spin, and constrains any Comptonized contamination through both its spectral and polarimetric signatures.

\subsubsection*{Models for coronal emission and disc reflection}

The first \texttt{KY} model to predict physically motivated polarization properties of the disc reflection was \texttt{KYNLPCR} \cite{Dovciak2011}, in contrast to the non-physical assumption of 100\% linear polarization perpendicular to the disc plane used in \texttt{KYNHREFL} (see Section~\ref{sec:spectral}). It computes the re-processed emission from a fully neutral accretion disc in the lamp-post geometry, where the primary source is isotropic and emits a cut-off power-law spectrum. The adopted lamp-post geometry is shown in the central panel of Figure \ref{fig:ky_geometries}, where the lamp-post at height $h$ above the black hole either directly emits towards the observer or illuminates the disc, which re-processes the primary emission before emitting towards the observer. The local disc re-processing is modelled by the Monte Carlo code \texttt{NOAR} \cite{Dumont2000, Rozanska2002}, which computes the reflection spectrum of a neutral disc and provides a slight limb-brightening emission directionality. The polarization is computed using the Rayleigh single-scattering approximation for both unpolarized and linearly polarized primary radiation \cite{Chandrasekhar1960}, with the fluorescent lines treated as intrinsically unpolarized. The relativistic rotation of the PA is accounted for along all photon paths — from the lamp-post to the disc, from the lamp-post directly to the observer, and from the disc to the observer. As with all \texttt{KY} package models, \texttt{KYNLPCR} includes additional parameters defining the emission region geometry and possible obscuration by a circular cloud, as described in Section~\ref{sec:ray-tracing}. The polarization timing properties of X-ray eclipses due to orbiting clouds over accretion discs were investigated in \cite{Kammoun2018} using the \texttt{KYNLPCR} model, demonstrating how the asymmetric obscuration of the relativistically distorted disc image introduces characteristic time-variable spectro-polarimetric signatures. While \texttt{KYNLPCR} was a pioneering tool for exploring GR effects on the polarization of reflected coronal emission, it is limited to a fully neutral disc and a single-scattering approximation for the polarization --- both significantly improved in the \texttt{KYNSTOKES} model described below.

\begin{figure}[t]
 \sidecaption
\includegraphics[scale=.22]{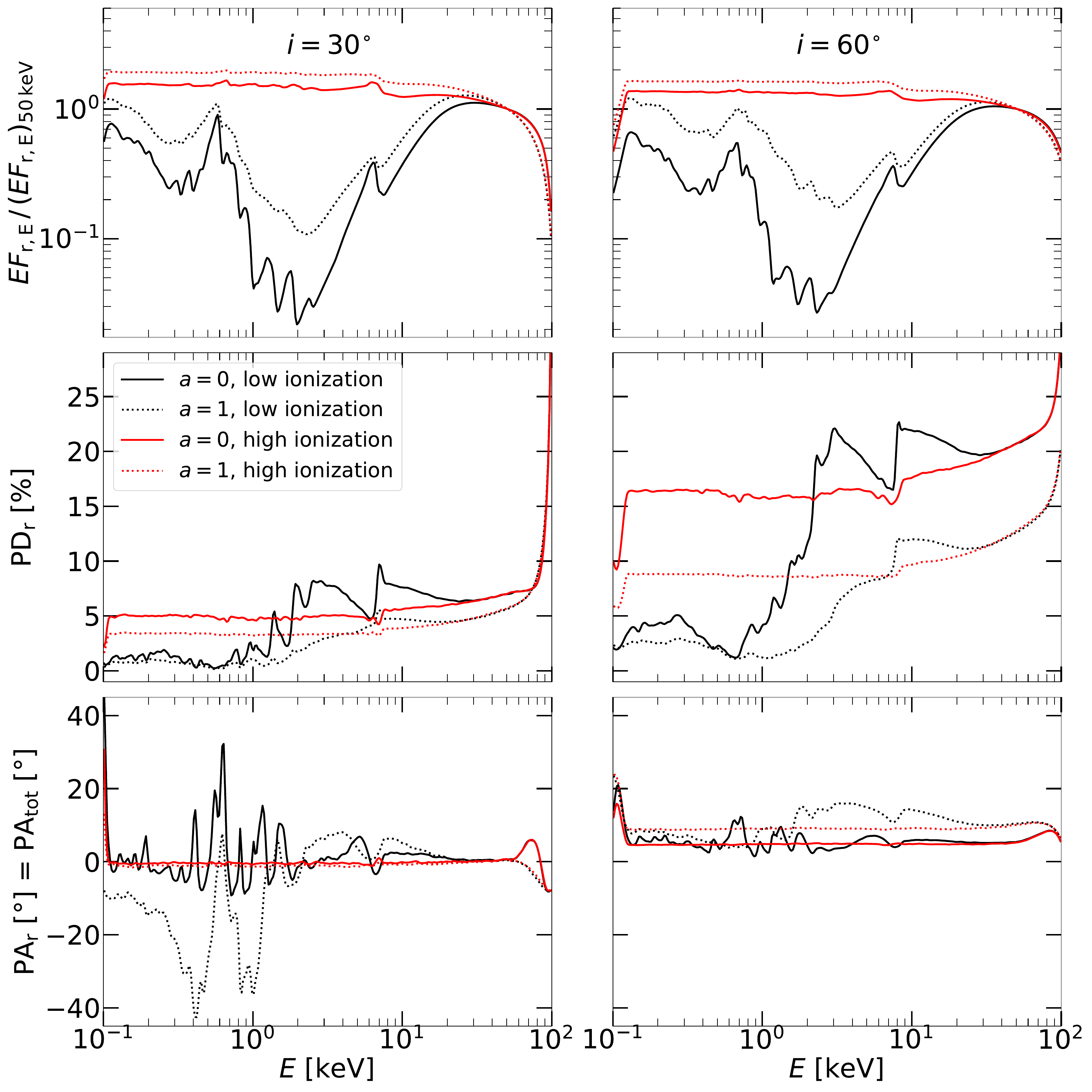}
%
\caption{\footnotesize Relativistically reflected X-ray emission from the AGN accretion disc extended to the ISCO, as predicted by the \texttt{KYNSTOKES} model for the isotropic unpolarized lamp-post corona (with $\mathrm{\Gamma} =2$ and $h = 3\,GM/c^2$) and a distant observer inclined at 30$^\circ$ (left column) and $60^\circ$ (right column). We show the renormalized spectra (top row), polarization degree (middle row), and polarization angle (bottom row). The results for the disc in low ionization (black) and high ionization (red) states, and for non-rotating BH ($a = 0$; solid) and maximally rotating BH ($a = 1$; dotted) are shown.
Image adapted from \cite{Podgorny2023a}.}
\label{fig:kynstokes_energydep}       
\end{figure}

A major step forward was the development of \texttt{KYNSTOKES} \cite{Podgorny2023a, Podgorny2023thesis}, which uses the \texttt{STOKES} partially ionized reflection tables described in Section~\ref{sec:1} for the local rest-frame reflection, replacing the neutral disc and single-scattering Rayleigh approximation of \texttt{KYNLPCR} with a full Compton multiple-scattering treatment of a partially ionized disc. \texttt{KYNSTOKES} operates in two geometries: the lamp-post and the slab corona co-planar with the disc. Returning radiation has not yet been implemented in either the lamp-post or slab corona geometry of \texttt{KYNSTOKES}. It is, however, partly addressed in the \texttt{KYNSTOKESRR} model described later, which includes the returning primary radiation in the slab corona geometry but not the secondary returning radiation of the once-reflected emission.

\begin{figure}[t]
 \sidecaption
\includegraphics[scale=.22]{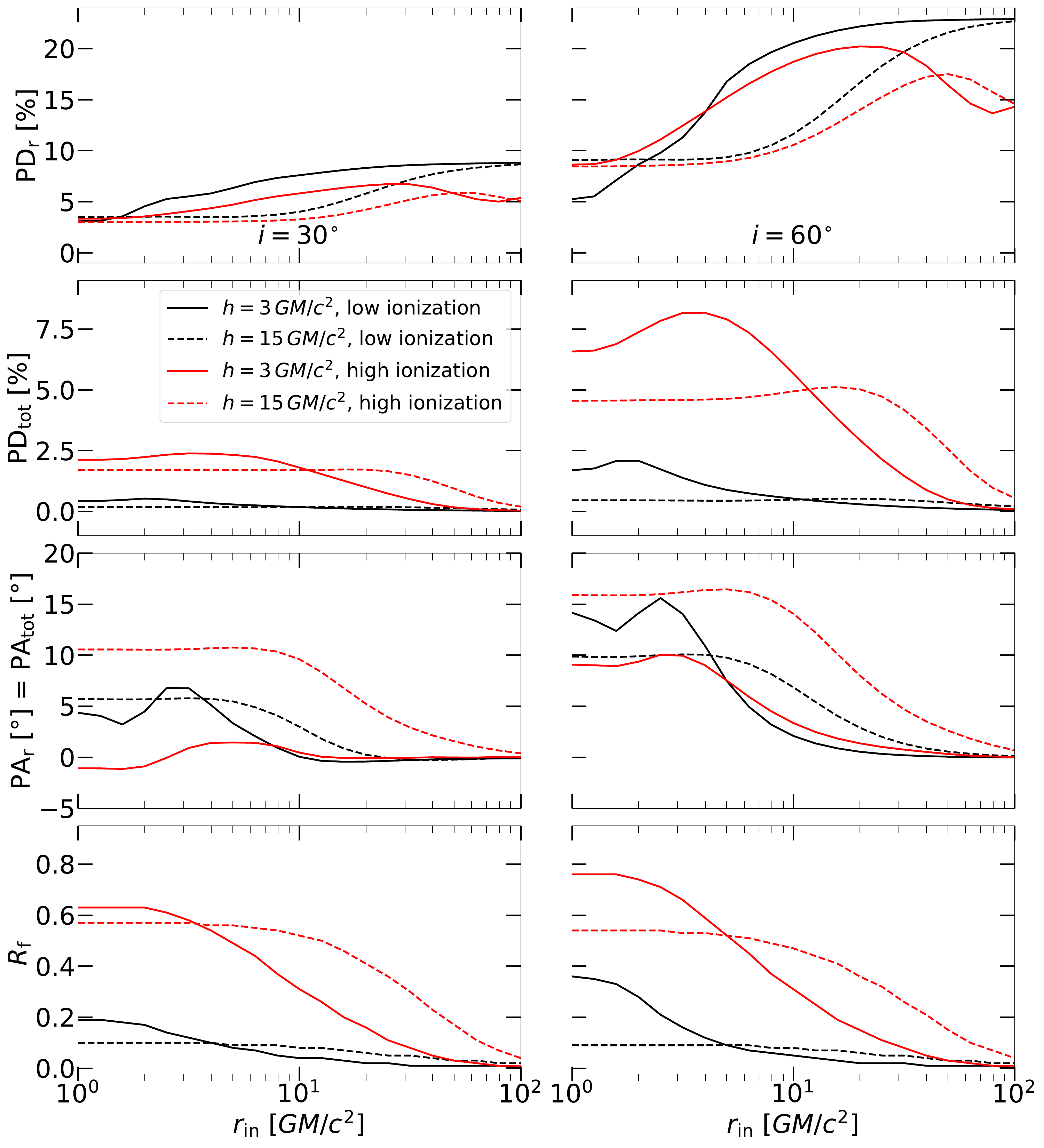}
%
\caption{\footnotesize The 2--8 keV integrated emission from the innermost accretion region of AGNs, as predicted by the \texttt{KYNSTOKES} model for the isotropic unpolarized lamp-post corona (with $\mathrm{\Gamma} =2$), maximally rotating BH ($a=1$), and a distant observer inclined at 30$^\circ$ (left column) and $60^\circ$ (right column). From top to bottom, we show the reflected-only polarization degree, the total (primary + reflected) polarization degree, the polarization angle (identical for the reflected-only and total emission since the primary source is unpolarized), and the reflected flux fraction from the total emission, all with respect to the position of the inner disc edge. 
The results for the disc in low ionization (black) and high ionization (red) states and for the lamp-post heights of $h=3\,GM/c^2$ (solid) and $h = 15\,GM/c^2$ (dashed) are shown.
Image adapted from \cite{Podgorny2023thesis}.}
\label{fig:kynstokes_rin}       
\end{figure}
Figure \ref{fig:kynstokes_energydep} shows the reflected-only spectra, PD and PA with energy for different ionization states of the disc extending to the ISCO, BH spins, and inclinations, as predicted by the \texttt{KYNSTOKES} model. The disc ionization state, BH spin, inclination, $r_\mathrm{in}$, and $h$ are all primary polarization drivers in the presumed global geometry. The energy dependence of the spectra and polarization on $\xi$ in the local co-moving frame (see Figure \ref{fig:results_loc_reflection_xi}) is recognizable here, relativistically smeared by the integration over the disc. The prevalent orientation of the reflection-induced PA is aligned with the projected system axis, because in the lamp-post geometry the dominant plane of scattering for disc reflection is close to equatorial, although the GR rotation of the PA causes a small observable deflection by $\lesssim 20^\circ$ from the projected axis direction, depending on the exact model configuration. X-ray polarization is sensitive to the disc truncation in relativistic reflection models, because the position of the inner disc edge determines the global geometry of scattering. Figure \ref{fig:kynstokes_rin} shows the reflected and total PD and PA, and the self-consistently computed flux reflection fraction, $R_\mathrm{f}$, in the 2--8 keV band with respect to $r_\mathrm{in}$, for two cases of disc ionization, $h$, and $i$, for a maximally rotating BH. Although the details depend on the exact configuration of the innermost accretion region, generally the more the disc is truncated at its inner edge, the more the geometry of scattering is reduced and the reflection-induced PD increases, to the detriment of the observed reflection fraction.

Contrary to the direct coronal emission, for relativistic disc reflection the lamp-post geometry produces a more asymmetric configuration than the slab geometry, inducing higher reflection-induced PD. This PD is comparable to, and can exceed, the values predicted by Comptonization in slab-like coronae, with a similar dependence of PD on inclination and similar average PA aligned with the projected axis.

\begin{figure}[b]
 \sidecaption
\includegraphics[scale=.40,
                   trim={0.0cm 0.0cm 17cm 0.0cm},
                   clip]{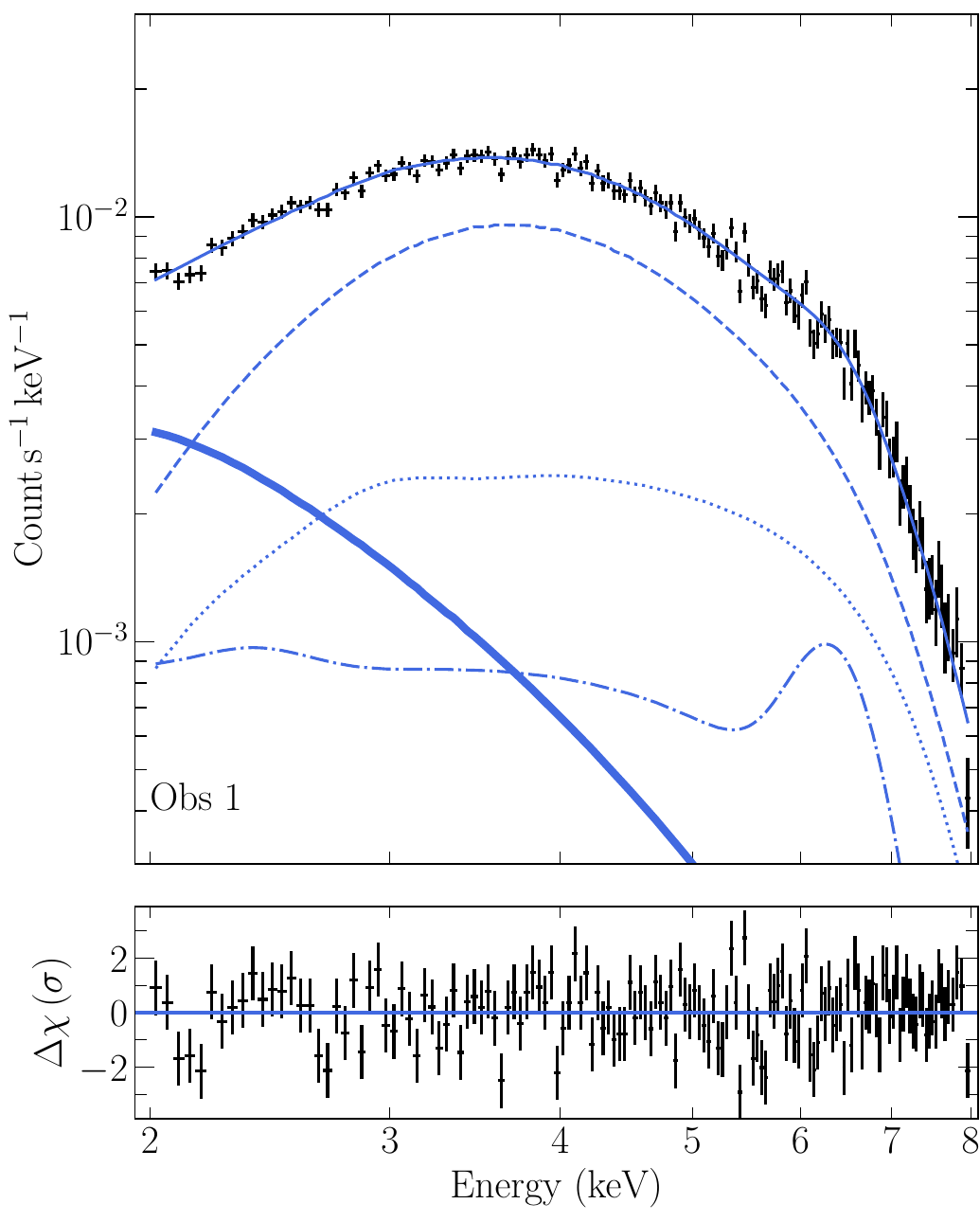}
\includegraphics[scale=.45,
                   trim={0.0cm 0.0cm 34.1cm 0.9cm},
                   clip]{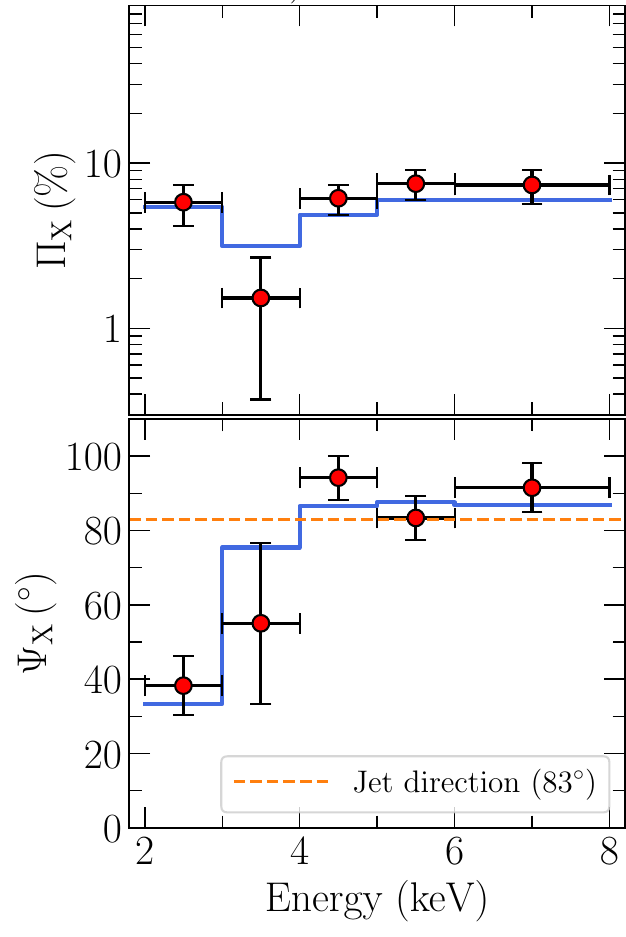}
%
\caption{\footnotesize Example of usage of the \texttt{KYNSTOKES} and \texttt{STOKES\_TORUS} models for spectro-polarimetric fitting of the NGC 4151 observation by {\it IXPE} in \texttt{XSPEC}. On the left, we show the fitted spectra and residuals for one of the two observations, including decomposition into individual components. The primary source (dashed) and relativistic disc reflection (dotted) are given by the \texttt{KYNSTOKES} model, the distant torus reflection (dot-dashed) is given by the \texttt{STOKES\_TORUS} model, and the soft-X-ray scattered component is given by a simple power-law model (thick solid). On the right, we show the simultaneously fitted polarization degree (top) and polarization angle (bottom) in 5 energy bins across the {\it IXPE} energy range for both observations combined. The orange line shows the projected direction of the radio jet of the system. Image adapted from \cite{Kammouninprep}.}
\label{fig:kynstokes_4151}       
\end{figure}
The \texttt{KYNSTOKES} model has been applied to both AGN and XRB systems observed by {\it IXPE}. Using the spectral parameters from the existing fit of the Cygnus X$-$1 XRB in the hard state, the \texttt{KYNSTOKES} model predictions in the lamp-post regime showed that the observed reflection fraction is too low to produce the high PD detected by {\it IXPE} \cite{Krawczynski2022} --- a larger reflection fraction would be needed to reach the required polarization level. On the other hand, for the case of the NGC 4151 AGN, the lamp-post coronal approximation may be a viable alternative to the equatorial coronal geometry 
when we take into account the high reflection-induced polarization. A fit with the \texttt{KYNSTOKES} model in the lamp-post geometry \cite{Kammouninprep} showed that the observed $\sim 6\%$ of PD with PA aligned with the system axis may be attributed primarily to the highly polarized relativistic disc reflection, assuming an unpolarized lamp-post illumination. Figure \ref{fig:kynstokes_4151} shows the resulting \texttt{XSPEC} spectro-polarimetric fit of the {\it IXPE} data. In addition to \texttt{KYNSTOKES}, which represents the inner disc-corona system, a scattered power-law component is needed to explain the different PA observed at the lowest energies, and the \texttt{STOKES\_TORUS} model is used to account for the distant nearly neutral reflection, needed primarily to fit the narrow iron line. 

\begin{figure}[t]
 \sidecaption
\includegraphics[scale=.38]{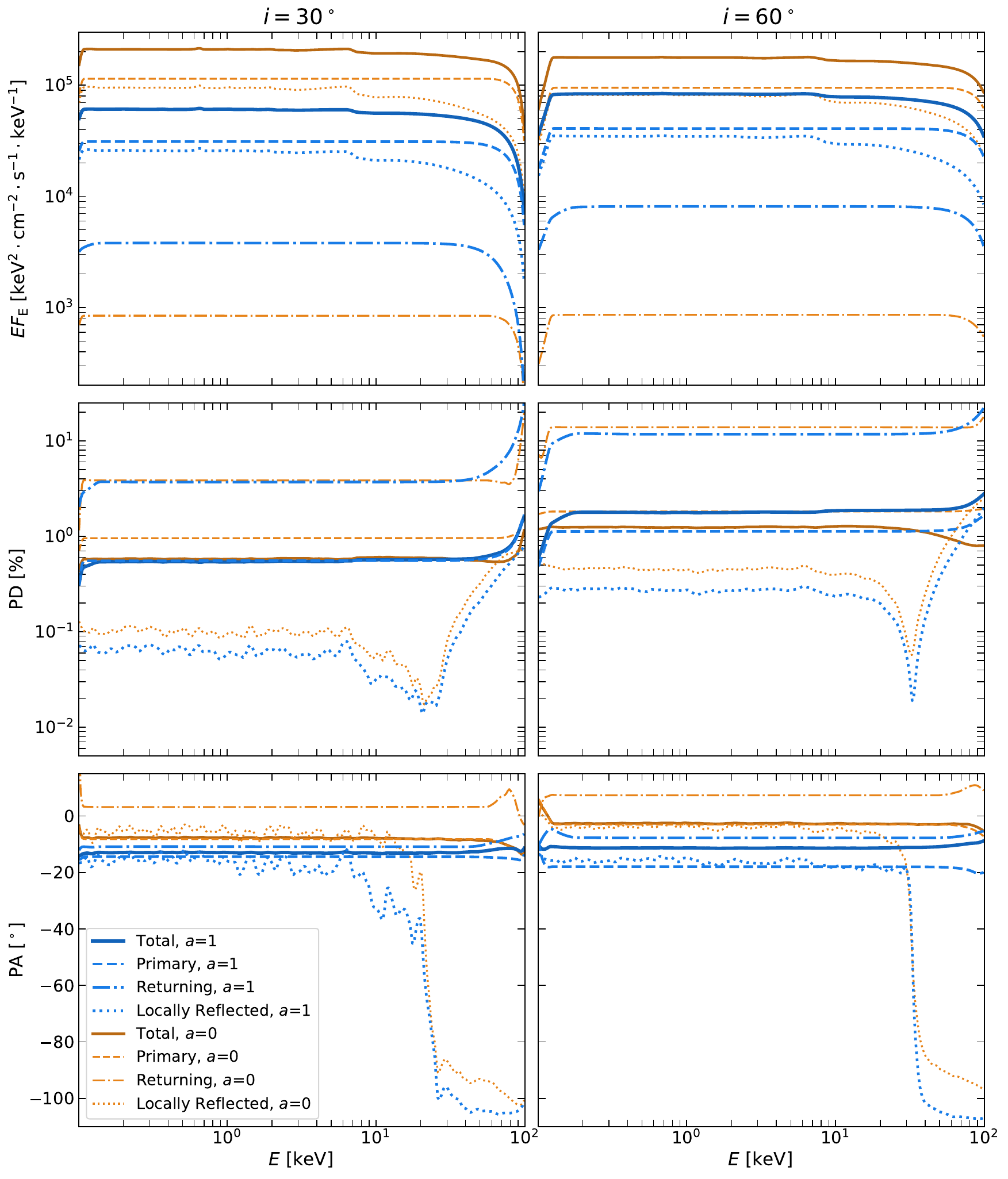}
%
\caption{\footnotesize The X-ray AGN emission from a system with slab corona above the accretion disc (with $\sim r^{-3}$ radial emissivity profile), as predicted by the \texttt{KYNSTOKESRR} code, which includes the primary emission (with $\mathrm{\Gamma} =2$ and polarized at 2\% parallel to the system axis), locally reflected emission from a highly ionized disc extending to the ISCO, 
and reflected returning coronal emission re-processed in the electron-scattering atmosphere with a 100\% albedo. We show the spectra (top row), polarization degree (middle row), and polarization angle (bottom row), for a distant observer inclined at $30^\circ$ (left column) and $60^\circ$ (right column). The results for maximally rotating ($a = 1$; blue) and non-rotating ($a = 0$; orange) BH are shown. Image courtesy of Matěj Ptáček.}
\label{fig:kynstokes_extendedRR}       
\end{figure}

Recently, \texttt{KYNSTOKES} was extended to include the reflection of returning primary radiation in the slab corona geometry shown in the right panel of Figure \ref{fig:ky_geometries}, giving rise to the \texttt{KYNSTOKESRR} model variant. Preliminary studies of the reflection polarization properties for the slab coronal geometry have been presented in \cite{Podgorny2023thesis, Ptacek2025thesis}, but a more detailed investigation still remains to be done. The local re-processing for isotropic power-law irradiation of the disc below the corona is computed with the \texttt{STOKES} tables and typically results in an observed PA alternating between close to parallel and perpendicular to the projected system axis in the X-ray band and very low PD of the component. The polarization of the reflection of returning radiation is currently implemented only with Chandrasekhar's formulae for diffuse reflection from a semi-infinite electron-scattering atmosphere with adjustable energy-independent albedo.
The model does not account for scattering of reflected photons within the corona, neither for the locally reflected emission escaping upward through the coronal layer, nor for the returning radiation before or after its reflection from the disc.
Results from \texttt{KYNSTOKESRR} are presented in Figure \ref{fig:kynstokes_extendedRR} for different BH spins and inclinations. Similarly to the studies of reflected returning thermal radiation, the reflected returning coronal radiation is shown to be strongly polarized with PA nearly parallel to the projected system axis. Therefore, the \texttt{KYNSTOKESRR} predictions independently support the conclusion of \cite{Steiner2024} that returning radiation plays a dominant role in the observed polarization properties of Cygnus X$-$1 in the soft state, as interpreted using the \texttt{KERRC} code \cite{Krawczynski2022b}.


\subsubsection*{Ongoing development}

Many of the \texttt{KY} codes suitable for X-ray polarization studies are under constant development. The current efforts mostly focus on implementing the \texttt{STOKES} power-law reflection tables into the \texttt{KYNSTOKESRR} model and the newest \texttt{STOKESBBTRANS} blackbody transmission tables into the \texttt{KYNBBRR} model. All variants of \texttt{STOKES} tables themselves are being updated to include Comptonization on hot electrons inside the disc atmosphere (Gupta et al., in prep., see \cite{Podgorny2025} for preliminary results and the {\tt STOKES} tables repository for version 2.1). Further awaited upgrades include relaxing the assumption of isotropic lamp-post emission and development of a warm Comptonization model. In the \nameref{technical_docs}, we provide references to the codes' repositories where the most recent public updates may be tracked until this chapter is updated in a future edition.


\begin{acknowledgement}
JP and MD would like to thank Matěj Ptáček for producing the sketches for this chapter and Figure \ref{fig:kynstokes_extendedRR}. JP and MD were supported by GACR project 26-22614S and acknowledge institutional support from RVO:67985815.
\end{acknowledgement}

\clearpage
\section*{Appendix}
\addcontentsline{toc}{section}{Appendix}
\phantomsection
\label{technical_docs}

Tables \ref{tab:models1}--\ref{tab:models4} list online repositories for the model tables and codes discussed in the chapter. If a public version of the model exists, its repository contains technical documentation, instructions for usage, a link for download, and a list of relevant references.

\begin{tabularx}{\linewidth}{@{} l @{\hspace{1em}} X @{}}
  \caption{{\footnotesize Overview of the reflection and transmission tables computed with  \texttt{STOKES} code}}
  \label{tab:models1}\\
  \toprule
  \textbf{\small Table} & \textbf{\small Repository} \\ 
   & \textbf{\small Notes} \\
  \midrule
  \endfirsthead
  \bottomrule
  \endfoot
  \bottomrule
  \endlastfoot
      \small \texttt{STOKES}
      & \small \url{https://github.com/jpodgorny/stokes\_tables}\\
      & \small Local reflection tables for an incident power-law radiation impinging on a partially ionized constant-density  plane-parallel slab atmosphere, as computed with the \texttt{TITAN} and \texttt{STOKES} codes in PIE, or a fully neutral constant-density slab atmosphere. They are in the FITS format of the \texttt{XSPEC} OGIP standard and can be used directly in \texttt{XSPEC} via the \texttt{atable} command.\\[0.5em]\hline
      \small \texttt{STOKESBB}
      & \small \url{https://github.com/jpodgorny/stokesBB\_tables}\\
      & \small Local reflection tables for an incident blackbody radiation impinging on a partially ionized constant-density plane-parallel slab atmosphere, as computed with the \texttt{TITAN} and \texttt{STOKES} codes in PIE, or a fully neutral constant-density slab atmosphere. They are in the FITS format of the \texttt{XSPEC} OGIP standard and can be used directly in \texttt{XSPEC} via the \texttt{atable} command.\\[0.5em]\hline
      \small \texttt{STOKESBBTRANS}
      & \small \url{https://github.com/jpodgorny/stokesBBtrans\_tables}\\
      & \small Local blackbody transmission tables for a plane-parallel atmosphere of finite optical depth. The repository currently contains only v1.0 tables in the FITS format for an incident blackbody radiation impinging on a fully ionized constant-density atmosphere, as computed with the \texttt{STOKES} code for various optical depths and Thomson scattering. The variants for partially ionized atmospheres and Compton scattering are not yet publicly available. \\[0.5em]\hline
\end{tabularx}

\clearpage
\begin{tabularx}{\linewidth}{@{} l @{\hspace{1em}} X @{}}
   \caption{{\footnotesize Overview of the \texttt{XSPEC} non-relativistic \texttt{STOKES} models for distant reflection}}
   \label{tab:models2}\\
  \toprule
  \textbf{\small Model} & \textbf{\small Repository} \\ 
   & \textbf{\small Notes} \\
  \midrule
  \endfirsthead
  \bottomrule
  \endfoot
  \bottomrule
  \endlastfoot
    \small \texttt{STOKES\_DISC} 
      & \small \url{https://github.com/jpodgorny/stokes\_disc}\\ 
      & \small Nearly neutral power-law reflection from a distant geometrically thin disc illuminated by an extended central source. \\[0.5em]\hline
    \small \texttt{STOKES\_CONE} 
      & \small \url{https://github.com/jpodgorny/stokes\_cone}\\
      & \small Power-law reflection from a distant vertically extended double-cone illuminated by a~central compact source. \\[0.5em]\hline
      \small \texttt{STOKESBB\_CONE} 
      & \small \url{https://github.com/jpodgorny/stokesbb\_cone} \\
      & \small Blackbody reflection from a distant vertically extended double-cone illuminated by a~central compact source. \\[0.5em]\hline
    \small \texttt{STOKES\_TORUS} 
      & \small \url{https://github.com/jpodgorny/stokes\_torus}\\ 
      & \small Power-law reflection from a distant vertically extended elliptical torus illuminated by a~central compact source. \\[0.5em]\hline
      \small \texttt{STOKESBB\_TORUS} 
      & \small \url{https://github.com/jpodgorny/stokesbb\_torus} \\ 
      & \small Blackbody reflection from a distant vertically extended elliptical torus illuminated by a~central compact source. \\[0.5em]\hline
    \small \texttt{STOKES\_BOWL} 
      & \small \url{https://github.com/jpodgorny/stokes\_bowl}\\
      & \small Power-law reflection from a distant vertically extended double-bowl illuminated by a~central compact source. \\[0.5em]\hline
      \small \texttt{STOKESBB\_BOWL} 
      & \small \url{https://github.com/jpodgorny/stokesbb\_bowl} \\
      & \small Blackbody reflection from a distant vertically extended double-bowl illuminated by a~central compact source. \\[0.5em]\hline
\end{tabularx}

\clearpage
\begin{tabularx}{\linewidth}{@{} l @{\hspace{1em}} X @{}}
   \caption{{\footnotesize Overview of the \texttt{XSPEC} relativistic spectral and timing \texttt{KY} models}}\label{tab:models3}\\
  \toprule
  \textbf{\small Model} & \textbf{\small Repository} \\ 
   & \textbf{\small Notes} \\
  \midrule
  \endfirsthead
  \bottomrule
  \endfoot
  \bottomrule
  \endlastfoot
    \parbox[t]{0.15\linewidth}{\small \texttt{KYNRLINE} \texttt{KYNRLPLI} \texttt{KYNCONV} \texttt{KYNCLP}}
      & \parbox[t]{\linewidth}{\small \url{https://projects.asu.cas.cz/stronggravity/kyn}\\[0.2em]
        Relativistic line profile and convolution models, including simple non-physical polarization implementation. Not recommended for quantitative spectro-polarimetric analysis.}\\[0.5em]\hline
    \parbox[t]{0.15\linewidth}{\small \texttt{KYNHREFL} \texttt{KYNREFIONX} \texttt{KYNXILLVER}}
      & \parbox[t]{\linewidth}{\small \url{https://projects.asu.cas.cz/stronggravity/kyn}\\[0.2em] 
      Relativistic disc reflection models. While \texttt{KYNHREFL} includes simple non-physical polarization implementation, it is not recommended for quantitative spectro-polarimetric analysis.}\\[0.5em]\hline
    \small \texttt{KYNSED} 
      & \small \url{https://projects.asu.cas.cz/dovciak/kynsed}\\ 
      & \small Relativistic spectral model for thermal disc emission and disc reflection in the lamp-post scenario with self-consistent disc-corona interaction, covering X-ray, UV, and optical bands.\\[0.5em]\hline
    \parbox[t]{0.15\linewidth}{\small \texttt{KYNREFREV} \texttt{KYNXILREV}}
      & \parbox[t]{\linewidth}{\small \url{https://projects.asu.cas.cz/stronggravity/kynreverb}\\[0.2em]
        \texttt{KYNREVERB} package of reflection-induced X-ray reverberation models in the lamp-post scenario, counterparts of \texttt{KYNREFIONX} and \texttt{KYNXILLVER}.}\\[0.5em]\hline
    \small \texttt{KYNXILTR} 
      & \small \url{https://projects.asu.cas.cz/dovciak/kynxiltr}\\ 
      & \small X-ray reflection-induced and UV/optical thermal disc reverberation model for AGN in lamp-post scenario with self-consistent disc-corona interaction, where the UV/optical reverberation arises from the thermal disc emission heated by the corona. \\[0.5em]\hline
    \small \texttt{KYNSPOT} 
      & \small Not yet publicly available.\\
      & \small Relativistic orbiting spot timing model (used outside of \texttt{XSPEC}). Not yet updated for the most recent format of \texttt{KY} codes.\\[0.5em]\hline
\end{tabularx}
  
\clearpage
\begin{tabularx}{\linewidth}{@{} l @{\hspace{1em}} X @{}}
   \caption{{\footnotesize Overview of the \texttt{XSPEC} relativistic spectro-polarimetric \texttt{KY} models}}\label{tab:models4}\\
  \toprule
  \textbf{\small Model} & \textbf{\small Repository} \\ 
   & \textbf{\small Notes} \\
  \midrule
  \endfirsthead
  \bottomrule
  \endfoot
  \bottomrule
  \endlastfoot
    \parbox[t]{0.15\linewidth}{\small\texttt{KYNPHEBB} \texttt{KYNBB}}
      & \parbox[t]{\linewidth}{\small \url{https://projects.asu.cas.cz/stronggravity/kyn}\\[0.2em]
        Relativistic spectro-polarimetric models for thermal disc emission with a phenomenological power-law profile of the disc temperature and the Novikov-Thorne disc (without the absorption effects and returning radiation). The \texttt{STOKESBBTRANS} tables (v1.0) for the direct emission are included.}\\[0.5em]\hline
      \parbox[t]{0.15\linewidth}{\small \texttt{KYNBBRR} \texttt{KYNFBBRR}}
      & \parbox[t]{\linewidth}{\small Not yet publicly available.\\[0.2em]
      Relativistic spectro-polarimetric models for thermal emission from the Novikov-Thorne disc, including returning radiation. The latter variant includes the \texttt{STOKESBBTRANS} tables (v1.0) for the direct emission and further geometrical parameters. The variants with more complex rest-frame atmospheric re-processing (with higher versions of the tables)
      are similarly not yet publicly available.}\\[0.5em]\hline
    \small \texttt{KYNLPCR}
      & \small \url{https://projects.asu.cas.cz/stronggravity/kyn}\\
      & \small Relativistic spectro-polarimetric model for the disc reflection of coronal emission in the lamp-post scenario and neutral disc atmosphere. Single scattering approximation is used for local polarization properties.\\[0.5em]\hline
    \small \texttt{KYNSTOKES} 
      & \small \url{https://projects.asu.cas.cz/dovciak/kynstokes}\\
      & \small Relativistic spectro-polarimetric model for the disc reflection of coronal emission in lamp-post and slab corona geometries with a pre-computed re-processing in partially ionized disc atmosphere.\\[0.5em]\hline
    \small \texttt{KYNSTOKESRR} 
      &  \small Not yet publicly available.\\
      & \small Relativistic spectro-polarimetric model for the disc reflection of coronal emission in lamp-post and slab corona geometries with a pre-computed re-processing in partially ionized disc atmosphere including reflection of returning coronal radiation in the slab corona geometry.\\[0.5em]\hline
\end{tabularx}

\clearpage
As an example of the application of the non-relativistic \texttt{STOKES} models for distant reflection, Table \ref{best-fit_Cyg_X-3} shows the best-fit parameter values for the model fit presented in Figure \ref{fig:Cyg_X-3_fit}, which shows the spectro-polarimetric fit of the 2022 and 2024 {\it IXPE} observations of Cygnus X$-$3 in the hard, intermediate, and soft states
using the \texttt{STOKES\_BOWL} and \texttt{STOKESBB\_BOWL} models.
The fit was performed with an added 1\% systematic error, as in \cite{Veledina2024}. The data were rebinned from the original 150 energy channels to 30 energy bins in 2--8 keV. The system orientation, $\mathrm{\Delta\Psi}$, which is driven by the measured PA, was first fitted preliminarily for the hard state and then kept frozen during the fitting and error estimation of the remaining parameters and states. 
The spectral hardness set by the incident power-law index $\mathrm{\Gamma}$ or blackbody energy $kT_\mathrm{BB}$ is degenerate with the absorption column density $N_\mathrm{H}$. The radial power-law index $\beta = -2$ of the ionization parameter implies that, for a given flux anisotropy, the reflector's density decreases as $\sim r^{-4}$, which is natural for super-Eddington winds. The model allows one to determine $\rho_\mathrm{in}$ from the fitted normalization and $\xi_0$ values for a consistency check, provided the expected neutral hydrogen density $n_\mathrm{H,0}$ at $\rho_\mathrm{in}$, the distance to the source $D$, and the BH mass $M$ are known. Assuming $n_\mathrm{H,0} = 10^{18} \, \mathrm{cm}^{-3}$, $D = 9.67$ kpc, and $M = 10\,M_\odot$, we obtain $\rho_\mathrm{in} = 351, 218, \textrm{and} \,251 \, GM/c^2$ for the hard, intermediate, and soft states.
\begin{table}
\footnotesize
\centering
\caption{{\footnotesize Best-fit parameter values with $1\sigma$ errors for the \texttt{CONST * TBABS * STOKES\_BOWL} model applied to the 2022 {\it IXPE} observations of Cygnus X$-$3 in the hard and intermediate states. And analogously, for the \texttt{CONST * TBABS * STOKESBB\_BOWL} model applied to the 2024 {\it IXPE} observations in the soft state. No redshift is applied.}}
\label{best-fit_Cyg_X-3}
\begin{tabular}{l@{\hspace{1.2cm}}c@{\hspace{1.2cm}}c@{\hspace{1.2cm}}c}
\hline
Parameter [unit] & Hard state & Intermediate state & Soft state \\
\hline\hline
$N_{\rm H}$ [$10^{22}\,{\rm cm}^{-2}$] & $10.08^{+0.07}_{-0.08}$ & $7.51^{+0.09}_{-0.13}$ & $3.65\pm0.06$\\
\hline
$\mathrm{\Gamma}$ & $2.59^{+0.02}_{-0.01}$ & $2.49^{+0.05}_{-0.06}$ & --- \\
$kT_\mathrm{BB}$ [keV] & --- & --- & $1.012^{+0.003}_{-0.001}$ \\
$\cos(i)$ & 0.866 (frozen) & 0.866 (frozen)  &  0.866 (frozen) \\
$\mathrm{\Theta}$ [$^\circ$] & $15.1^{+0.5}_{-0.3}$ & $9.6^{+0.6}_{-0.7}$ & $5.7\pm0.1$\\
$\rho/\rho_\mathrm{in}$ & $1.9^{+0.2}_{-0.1}$ & $1.2^{+0.4}_{-0.1}$ & $1.201^{+0.001}_{-0.005}$ \\
$\xi_0$ [${\rm erg\,\,cm\,\,s^{-1}}$] & $101.4^{+0.3}_{-1.0}$ & $747^{+199}_{-181}$ & $998^{+16}_{-11}$ \\
$\beta$ & -2 (frozen) & -2 (frozen) & -2 (frozen) \\
$\mathrm{PD}_\mathrm{0}$ & -1 (frozen) \textsuperscript{(a)} & 0.04 (frozen) \textsuperscript{(b)} & 0 (frozen) \textsuperscript{(b)} \\
$\mathrm{PA}_\mathrm{0}$ [$^\circ$] & --- & 0 (frozen) & --- \\
$\mathrm{\Delta\Psi}$ [$^\circ$] & 0.15 (frozen) & 0.15 (frozen) & 0.15 (frozen) \\
\texttt{norm} $[10^{-9}]$ & $3.0^{+0.4}_{-0.2}$ & $1.8^{+0.4}_{-0.2}$ & $1.5^{+0.3}_{-0.2}$ \\
\hline
const DU 1 & 1.0 (frozen) & 1.0 (frozen) & 1.0 (frozen) \\
const DU 2 & $1.034\pm0.004$ & $1.040\pm0.004$ & $1.035\pm0.004$ \\
const DU 3 & $1.029\pm0.004$ & $1.028\pm0.004$  & $1.029\pm0.004$  \\
\hline\hline
$\chi^2/\textrm{d.o.f.}$ & 383/234 & 241/234 & 273/234 \\
\hline
\end{tabular}
\vspace{0.5ex}
\begin{flushleft}
\footnotesize
\textsuperscript{a} Means that the intrinsic flux anisotropy and polarization distribution is according to a typical slab-like corona (see \cite{Podgorny2025b} for details). \\
\textsuperscript{b} Assumes isotropic intrinsic emission (see \cite{Podgorny2025b} for details), which may better correspond to neglected multiple reflections within the cavity in the model; an effect more important for higher ionization.
\end{flushleft}
\end{table}
\normalsize




\bibliographystyle{utphys-forcejournal-astro.bst}
\bibliography{ref.bib}








\end{document}